# Deep U band and R imaging of GOODS-South: Observations, data reduction and first results. [1] [2] [3]

Nonino, M.[1], Dickinson, M.[2], Rosati, P.[3], Grazian, A.[4], Reddy, N.[2], Cristiani, S.[1], Giavalisco, M.[5], Kuntschner, H.[6], Vanzella, E.[1], Daddi, E.[7], Fosbury, R. A. E.[6], Cesarsky, C.[8]

## ABSTRACT

We present deep imaging in the $U$ band covering an area of 630 arcmin$^2$ centered on the southern field of the Great Observatories Origins Deep Survey (GOODS). The data were obtained with the VIMOS instrument at the ESO Very Large Telescope. The final images reach a magnitude limit $U_{lim} \approx 29.8$ (AB, $1\sigma$, in a $1''$ radius aperture), and have good image quality, with full width at half maximum $\approx 0.8''$. They are significantly deeper than previous U–band images available for the GOODS fields, and better match the sensitivity of other multi–wavelength GOODS photometry. The deeper U–band data yield significantly improved photometric redshifts, especially in key redshift ranges such as $2 < z < 4$, and deeper color–selected galaxy samples, e.g., Lyman–break galaxies at $z \approx 3$. We also present the coaddition of archival ESO VIMOS R band data, with $R_{lim} \approx 29$ (AB, $1\sigma$, $1''$ radius aperture), and image quality $\approx 0.75''$. We discuss the strategies for the observations and data reduction, and present the first results from the analysis of the coadded images.

June 23, 2009

*Subject headings:* galaxies: evolution; galaxies: high-redshift

[1]INAF- Osservatorio Astronomico di Trieste , Via Tiepolo 11, I-34131 Trieste, Italy, e-mail: nonino@oats.inaf.it

[2]NOAO, 950 N. Cherry Avenue, Tucson, AZ 85719, USA

[3]European Southern Observatory, Karl-Schwarzschild-Str. 2, D-85748 Garching, Germany

[4]INAF - Osservatorio Astronomico di Roma, via Frascati 33, 00040, Monte Porzio Catone, Italy

[5]Department of Astronomy, University of Massachusetts, Amherst, MA 01003, USA

[6]ST-ECF, Karl-Schwarzschild Str. 2, 85748 Garching, Germany

[7]CEA, Laboratoire AIM, Irfu/SAp, F-91191 Gif-sur-Yvette, France

[8]CEA Saclay, Haute-commissaire a l'Energie Atomique, F-91191 Gif-sur-Yvette, France

## 1. Introduction

A variety of different approaches have been developed to identify samples of high-redshift galaxies. Among them, surveys of Lyman Break Galaxies (LBG) (Giavalisco 2002; Steidel et al. 2004; Reddy et al. 2008; Reddy & Steidel 2009) have yielded the largest spectroscopically confirmed samples. The LBG method selects galaxies with bright ultraviolet (UV) continuum emission arising from relatively unobscured, active star formation. Other techniques, primarily based on near- or far-infrared emission, have also been used to identify populations of high redshift objects, including distant red galaxies (Franx et al. 2003), extremely red objects (McCarthy 2004; Daddi et al. 2000), "BzK" galaxies (Daddi et al. 2004), and submillimeter galaxies (Chapman et al. 2005; Smail et al. 2004). Galaxies with spectral energy distribution (SED) dominated by evolved stellar populations, or by young but heavily obscured stars, may have UV rest frame colors that place them outside the selection region in the color plane defined for the LBG, or may be simply too faint at optical wavelengths to be identified at all.

A significant amount of work has been put into exploring the intersection between various color-selected galaxy populations (e.g., Reddy et al. 2005; van Dokkum et al. 2006; Chapman et al. 2005) and their relative contribution , (e.g., Grazian et al. 2007) to global properties e.g., Mass Luminosity Function, Star Formaty History. Very deep rest-frame UV data can be helpful to address these issues.

Moreover, even when rest-frame UV selection techniques are adopted, the resulting statistical description of the parent population can be uncertain. For example, Steidel et al. (2004) and Le Fevre et al. (2005) found major discrepancies in the bright end of the UV Luminosity Function (LF) at $z \approx 3$. Other studies (Reddy et al. 2008; Yoshida et al. 2006; Bouwens et al. 2007) suggest an evolution mainly limited to the bright part of the LF ($L \geq L^*$), while Iwata et al. (2007); Sawicki & Thompson (2006) find the evolution occurring at the faint end ($L \leq L^*$). Part of these discrepancies can be ascribed to the details of the selection (e.g. inclusion of AGNs, Reddy et al. 2008) and prevent a robust comparison with theoretical models (e.g. Marchesini & van Dokkum 2007). The implications of these differences for the history of star formation in galaxies are also very different. If the faint end slope of the LF is steep, as found by Steidel et al. (1999) and Reddy et al. (2008) at $z \approx 2$–3 and Bouwens et al. (2007) at $4 \leq z \leq 6$, then a substantial fraction of the UV luminosity density, $\rho_L = \int_0^{+\infty} L\phi(L)dL$, and thus the globally averaged rate of star formation in galaxies, arises from faint galaxies. If, instead, the faint end of the luminosity function is flatter (Gabasch et al. 2004; Sawicki & Thompson 2006) then the integrated UV luminosity density would be significantly lower, as well as the contribution of faint galaxies to the ionizing background. The contribution of LBG to the total stellar mass density of the Universe, e.g., Grazian et al. (2007), and to the X-ray number counts (Brandt & Hasinger 2005) critically depends on a robust determination of their LF. The uncertainties in the LF also affect the analysis of the clustering properties of LBG and their host dark matter halos (Giavalisco & Dickinson 2001; Lee et al. 2006, 2008).

One of the main aims of the Great Observatories Origins Deep Survey (GOODS, Giavalisco et al. (2004)) is the study of the formation and evolution of normal galaxies over a large range in redshift and stellar mass. However, the lack of deep U band imaging in both GOODS fields has so far limited our knowledge to the bright end of the luminosity function at $2 \leq z \leq 4$. .

In the GOODS-South field, previous U band observations obtained with the CTIO MOSAIC II (Dahlen et al. 2007) reach a limit of 26.7 AB mag ($5\sigma$). Very deep HST F300W observations, 27.5 AB mag ($10\sigma$), (de Mello et al. 2006a) cover only a very small fraction of the GOODS-South, and a shallower ($\approx 24.5 AB, 10\sigma$) but larger areal coverage has also been obtained (de Mello et al. 2006b). The ESO 2.2m WFI images of GOODS-South in $U_{38}$, B and R bands (Hildebrandt et al. 2005) are also relatively shallow,[12] reaching $5\sigma$ AB limits are 25.95, 27.35 and 27.15 (Hildebrandt et al. 2005, 2007; Taylor et al. 2009), respectively. The limitations of the U band data are particularly severe, as they are barely sufficient for robustly selecting $L^\star$ LBG at $z \approx 3$, which have $M^\star_{1700} = -20.97$, or $R_{AB} \approx 24.6$ (Reddy & Steidel 2009), and correspondingly much fainter U band magnitudes.

Smaller photometric errors improve the robustness of high redshift Lyman break selection by reducing the loss of faint galaxies from the LBG color selection window and minimizing contamination by lower-redshift interlopers. This in turn improves the dynamic range in luminosity for investigating galaxy properties at $z \approx 3$. Ultimately, by extending $U$–dropout LBG selection to fainter magnitudes, we aim to match the depth and dynamic range of other GOODS multi band data, e.g., the ultra-deep GOODS IRAC data suitable for measuring stellar masses, or the deep ACS imaging suitable for analysis of galaxy morphologies and for color selection of LBG at still higher redshifts.

---

[12]Observations of GOODS-South using the WFI $U_{50}$ filter could be affected by a red leak; see http://www.eso.info/sci/facilities/lasilla/instruments/wfi/inst/filters.

To reach this goal, we have carried out a campaign of deep U band imaging with VIMOS (Le Fevre et al. 2003) at the ESO VLT. In this paper, we present the final coadded image, as well as and initial results on LBG color selection and photometric redshifts using the new VIMOS U (hereafter $U_V$) band data. We also release a new deep and well-calibrated VIMOS R (hereafter $R_V$) band image, constructed by combining data from a number of archival programs, which is also useful for the selection of LBG at $z \approx 3$. These science-ready images are being released to the community via the ESO Archive. [13]

This paper is organized as follows. The strategy for the $U_V$ band observations, the data set itself, and the data reduction procedures are presented in §2. In §3, we describe the simulations carried out to characterize the released datasets, while in §4 the first results on LBG selected from the coadded images are presented. In §5 we summarize the conclusions.

Throughout this paper a $\Lambda-CDM$ concordance cosmological model with $H_0, \Omega_{\text{tot}}, \Omega_m, \Omega_\Lambda$ = 70 km/s/Mpc, 1.0, 0.3, 0.7 is adopted. Magnitudes are given in the AB system (unless otherwise stated).

## 2. Observations

The Great Observatories Origins Deep Survey (Giavalisco et al. 2004) covers two fields, one in the North centered on the Hubble Deep Field North (HDF-N) (Williams et al. 1996), and one in the South centered on the Chandra Deep Field South (Giacconi et al. 2002). The GOODS HST Treasury Program (Giavalisco et al. 2004) used the Advanced Camera for Surveys (ACS) to image both fields in four bands, F435W (hereafter $B_{435}$), F606W (V), F775W(i) and F850LP(z), reaching extended-source sensitivity similar to WFPC2 HDF observations. Over the past ten years a vast amount of data have been collected in the two fields, resulting in an unprecedented deep-multiwavelength coverage (see e.g., *http://www.stsci.edu/science/goods/*).

ESO has carried out several major observing campaigns to complement the HST and Spitzer GOODS datasets, including extensive spectroscopy (Vanzella et al. 2008, 2009;

---

[13]http://archive.eso.org/cms/eso-data/data-packages

Popesso et al. 2009), and near-infrared imaging (Retzlaff et al. 2009). As part of this effort, a program of deep $U_V$ band imaging was conducted with VIMOS at the UT3 in Service Mode, with a total time allocation of 40 hrs.

Used in imaging mode, VIMOS observes $2 \times 2$ fields of view, each using an EEV $2K \times 2K$ CCD detector that covers a field of view $\approx 7' \times 8'$ on the sky per quadrant, with gaps of $\approx 2'$ between the four fields. Each quadrant is equipped with its own filter. No atmospheric dispersion corrector is available, so observations are generally taken close to the meridian whenever possible. In order to fill the gaps between the four chips and cover the GOODS ACS field as uniformly as possible, a strategy was adopted using eight widely separated pointings, oriented with a position angle of $-20°$. Within each of these pointings, a series of 1000 s exposures was collected using a $20''$ dithering pattern. After rejecting 30 low quality single images, the final data set consists of 552 single chip images. We summarize the observing conditions in Table 1 and show an exposure map of the $U_V$ band observations in Figure 1.

The $R_V$ band data set was mainly constructed from the Large Programme 167.D-0492 in which the GOODS-South was repeatedly observed for a supernova search program. These observations consist of repeated individual exposures of 480 sec each mostly in only two pointings in the sky, with Position Angles -26° and -64°, therefore resulting in a mosaic with a pronounced gap roughly at the center of the area covered by ACS. In order to fill the gap, we used a large number of $R_V$ band exposures from the ESO Archive, mostly obtained as pre-imaging for spectroscopic programs, with exposure times ranging from 150 to 530 sec and different Position Angles. With these additional data, the $R_V$ band image consists of 610 single chip images. Figure 2 shows the color coded final coadded exposure map.

## 2.1. Data Reduction

Figure 3 schematically illustrates the procedure used to reduce the VIMOS imaging data. As a first step, bias images were subtracted and flat–field corrections were applied, using master flats constructed from twilight sky exposures. This provides a first gain correction of the four VIMOS chips. Using the bias and flats we also constructed a set of mask images, one per chip, to flag static bad pixels to be ignored in the following steps. After the bias and flat corrections, additional masks were created for satellite tracks, heavily vignetted regions, and other gross defects, by visual examination of each input image. Using Weight Watcher and SExtractor (Bertin & Arnouts 1996) with a relatively high detection threshold, a weight map was created for each image, in which detected cosmic rays and other artifacts were assigned zero value.

A source catalog was also produced and matched against the same astrometric catalog originally used to provide an astrometric grid for the ACS mosaic (Giavalisco et al. 2004) (see appendix A for details). After establishing a second order polynomial astrometric solution (see Appendix A), a first weighted coaddition of all 552 images was performed after subtracting the SExtractor-estimated background from each image. To perform the coaddition, the distortion coefficients derived from the astrometric solution were mapped into the Simple Image Polynomial (SIP) convention (Shupe et al. 2005, see also Appendix A). The weighted coaddition was carried out with the IRAF [14] task *wdrizzle* (Fruchter & Hook 2002), using a *Lanczos3* kernel.

In the coadding process, a major challenge was posed by the presence of diffuse, low-level light with a varying pattern in a significant number of images (Figure 4). To handle this effect we used a wavelet transform (WT) technique to estimate and subtract the background before the final coaddition. First, the preliminary stacked image was used to create a segmentation mask, running SExtractor with a relatively low detection threshold and then expanding the area covered by each detected object by 20%. Using the astrometric solution previously obtained and the segmentation mask from the first coaddition, we found in each single image the pixels corresponding to the detected objects in the coadded stack and replaced them with a random value from a Gaussian distribution with mean and standard deviation derived from a $\approx 30'' \times 30''$ box surrounding each pixel. In this way it was possible to avoid spurious effects due to pixels contaminated by objects, cosmic rays hits and other masked defects. The resulting images were wavelet transformed, using a six-level undecimated decomposition with an order 3 B-spline for the scaling function (Starck et al. 2007) derived from the MIDAS (Banse et al. 1983) code (see also Appendix D). The background of each frame was estimated using the lowest order plane of the WT.

In order to check whether the process of replacing masked objects with the surrounding background introduces biases into the large scale WT which would affect the photometry, we created a super-sky image from a median combination of 12 consecutive R band images (Figure 4). We then placed a grid of apertures replacing pixel values inside as described above, and computed the difference, within the apertures, between the lowest order WT of the original super-sky and the lowest order WT of the super-sky with pixels replaced. The standard deviation of the distribution of the differences ($\sigma_{diff}$) was then used with the expressions

---

[14]IRAF is distributed by the National Optical Astronomy Observatories, which is operated by the Association of Universities for Research in Astronomy, Inc., under cooperative agreement with the National Science Foundation.

$$\Delta mag = 1.086 * 3 * \frac{\sigma_{diff}}{source\_counts} \quad (1)$$

to derive $source\_counts$ from $\Delta mag$, and

$$mag = ZP - 2.5 \times log_{10}(source\_counts) \quad (2)$$

where ZP is the zero point of the mosaic image (see below) to find the magnitude of the source affected by the systematic $\Delta mag$. We found a bias which is negligible at bright magnitudes and which can reach a value of 0.15 mag (at $3\sigma$) for sources as faint as $R_{AB} = 27.5$. We then repeated the test for the $U_V$ band, which resulted in a possible 0.15 mag bias (at $3\sigma$) for sources as faint as $U_{AB} = 28.2$ and becoming progressively smaller for brighter objects. The underlying assumption of this procedure is that the diffuse light was an additive effect and not a multiplicative one. However, given the small amplitude of the scattered light, this assumption is not particularly important. Simply using the lowest order plane of the wavelet transform as a super-sky flat would change the photometry by less than 2%. In the final coaddition process we had to take into account both the remaining chip-to-chip gain variations and the changing photometric conditions among frames. The relative photometry was monitored with bright point sources measured with SExtractor separately in each chip for the all the exposures in a given pointing, after correcting for astrometric distortions. The photometric scaling derived in this way was used to create 32 coadded tiles (8 pointings with 4 chips each) for the $U_V$ band data and 8 ($2\times4$, data from 167.D-0492 only) for the $R_V$ band data. SExtractor was then again used perform tile-to-tile relative photometry, using the approach described in Koranyi et al. (1998), which minimizes the sum over all the photometric offsets among the tiles (see also Appendix B). A key assumption in the tile-to-tile photometric rescaling process is that the difference in the detector+filter responses in each VIMOS chip are small enough so that differences in color term negligible. This was effectively verified from the analysis of the photometric standards used to measure the zero points (see below). The final mosaic made from the weighted average coaddition of all the single processed frames was obtained by assigning to each single image a weight equal to $(rms^2 \times fluxscale \times FWHM^2)^{-1}$, where rms is the background rms of the input image, fluxscale is the photometric scaling factor and FWHM is the seeing of each image. In Figure 5 and Figure 6 we show the cumulative exposure time for the $U_V$ band and $R_V$ band mosaics in the area covered by GOODS-ACS and in the full areas respectively.

## 2.2. Photometric calibration

### 2.2.1. $U_V$ band

The zero point (ZP) of the coadded $U_V$ band image was derived from data taken on nights which were judged to be photometric from both the relative photometric analysis and from observations of standard star fields. These nights are marked with an asterisk (*) in Table 1. Figure 8 is a diagnostic diagram of the photometric calibration described below showing the difference between tabulated magnitude of Landolt (1992) stars, and those obtained for $U_V$ band as a function of magnitude and $(U - B)_{Landolt}$ color.

It should be noted that the $U_V$ band filter is significantly different for the Johnson U band filter (Bessel 1990) on which the Landolt system is based, and is also different from other U filters that have been used for Lyman break color selection. This is illustrated in Figure 7, where the $U_V$ band system transmission (including the filter, telescope and instrument optics, and the detector response) is taken from the VIMOS Exposure Time Calculator (ETC).[15] We established the $U_V$ band photometric system by requiring that the $U_V$ band magnitudes be equal to the Landolt U band magnitude for stars with zero colors in the Vega system. Since no stars with zero color were observed the zero point was obtained via least square fitting using the following expression:

$$ZP = 2.5 \log(\frac{\text{counts}}{\text{exptime}}) + m_{Landolt} - a_U \cdot \chi + a_1(U - B) + a_2(U - B)^2 + a_{B-V} * \chi \quad (3)$$

where $counts$ are the net counts of the standard star within an aperture of $7''$ radius, $m_{Landolt}$ is the Landolt U band magnitude, $\chi$ the airmass, and $a_U$ is the extinction coefficient (e.g., Da Costa 1992). The two color terms were introduced to account for the difference between Landolt U band filter and $U_V$ band filter. The values for $a_1$ and $a_2$ were found to be the same for all four chips within the uncertainties.

We further minimized the residuals by introducing a color dependent extinction term $a_{B-V}$ (for a similar approach see Holtzman et al. 1995). This was determined computing the color extinction of stars from the Pickles library (Pickles 1998) as a function of their B-V color, which is less sensitive than U-B to the effect of Balmer lines, by assuming the UV extinction curve used in the VIMOS ETC. We assumed that both the color terms and the color extinction term remained constant over the time spanned by the observations; their values are reported in Table 2. Since most of the observations of the standard stars did not span enough airmass range to properly determine the extinction coefficient, we used a value of $a_U = 0.45$, which was derived from the relative photometric analysis of data from photometric nights, and is in good agreement with the value of 0.42 determined for a single night when several standards were observed at varying airmasses.

---

[15]http://www.eso.org/observing/etc/bin/gen/form?INS.NAME=VIMOS+INS.MODE=imaging

The ZP of the final mosaic was obtained by matching the magnitude of bright sources in single images taken on photometric nights with the corresponding sources in the final coadded image. Aperture magnitudes measured within a $1''$ radius and corrected to total flux within a radius of $5''$ were used in order to take seeing differences into account. In Figure 9, (bottom panel) we show the residuals from this photometric matching.

The ZP thus obtained is 25.643 in the Vega system, which can be converted into the AB system with $U_V$ (AB) = $U_V$ (Vega) + 0.515[16]. This zero point has *not* been corrected for galactic extinction, which is E(B-V) = 0.008 at the position of GOODS-South (Schlegel 1998), as from NED [17]. The Cardelli et al. (1988) extinction relation would give $A_U = 0.04$. Given this small value, in the following we will use AB magnitudes with no galactic extinction correction.

As an independent check of the photometric ZP, we compared aperture-corrected photometry of bright sources with that derived from the CTIO U band image. These agree quite favorably (see top panel of Figure 9). In addition, the CTIO U band imaging was used to check for possible spatial variations in the relative ZPs, e.g., due to any illumination effect. We compared aperture photometry of cross-matched sources with $U_V \leq 20.7$ in the 32 coadded tiles described above. Photometric differences are plotted in Figure 10 against x and y position in each tile, and fitted with a second order polynomial. We find that the maximum residuals are no larger than 0.02 mag at the edges in both x and y, and thus do not apply any further corrections.

The spectrophotometric standards Feige 110 and SA98-193 were used to analyze the system response in the $U_V$ band. Both stars were observed on photometric nights and the observed $U_V$ band magnitudes were compared with the synthetic magnitude obtained by convolving the SED of Feige 110 (Bohlin et al. 2001) and of SA98-193 (Stritzinger et al. 2005) [18] with the $U_V$ band system response curve as given by the ESO ETC. For Feige 110 we measured a $U_V$ band magnitude of $10.64 \pm 0.01$ and a synthetic value of 10.619, while for SA98-193 we measured a magnitude of $12.39 \pm 0.035$ and a synthetic value of 12.436. These

---

stars have very different spectral types (DOp and K1III respectively, Drilling & Landolt 1979) and colors, so the difference between the magnitude offsets ($0.07 \pm 0.04$ mag) gives some indication of the degree to which uncertainties in the calibration of the $U_V$ bandpass may affect synthetic photometry.

### 2.2.2. $U_V$ filter red–leak

One of the four $U_V$ band filters is affected by a small red–leak at $\approx 4850$Å (Mieske et al. 2007). We found a value of $\approx 0.1\%$, similar to that reported in Mieske et al. (2007) from the spectra of the spectrophotometric standards LTT 7379 and Hiltner 600 with the $U_V$ band filter inserted. The effect of this read–leak on the selection of LBG is examined in Appendix C.

### 2.2.3. $R_V$ band calibration

For the $R_V$ band data, images from program 167.D-0492 observed on the night 2004:11:14 were calibrated using observations of Landolt standard star fields and including only the extinction term in eq. (1). The data from the other nights or other programs were anchored to the photometry from that one night, which was also used to set the mosaic ZP, 27.49 (AB). As an external check, we compare the aperture corrected photometry from the $R_V$ band mosaic and matched sources from the FORS1 R band data (Giacconi et al. 2002). The results of the comparison are reported in Figure 11.

## 2.3. Astrometry

To check the mosaic astrometry we matched the position of sources with magnitude between 18 and 24 from the $U_V$ band mosaic and the reference catalog and report the results in Figure 12. The matching radius was set to $1''$ and after 2 iterations with $3\sigma$ clipping, we obtained a standard deviation in Right Ascension and Declinations of approximately 66 mas, with no significant net coordinate offset. We also matched catalogs independently extracted from band $U_V$ and band R mosaics, matching radius $1''$ and using only objects with magnitude in the range 18-25 for $U_V$ and in the range 19-25 for R. The differences in Right Ascension and Declination are plotted in Figure 13.

### 2.4. Noise correlation, detection limits and completeness

To estimate the sky noise and thus the depth of the final images, we used SExtractor to create a segmentation map for each mosaic, which was then used to mask all the detected sources. We randomly placed apertures with radius $1''$ in regions of the $U_V$ coadded image with an effective exposure of 10 hrs or more, and measured counts using the the the IRAF *apphot.phot* task. After rejecting all apertures that encompass detected objects, a Gaussian fit to the distribution of measured counts within the $1''$ radius apertures gives a $1\sigma$ magnitude limit $U_V \approx 29.78$ AB mag.

For the completeness and detection limit analysis we used Skymaker (Erben et al. 2001) to generate images of artificial objects and add them to the $U_V$ band mosaic image. We randomly placed simulated point sources within the image, with FWHM ranging from $0.8''$to $2.0''$, and spanning a broad range of magnitudes. SExtractor was then used to detect sources in the image and the resultant catalog was matched (within a radius of 2 pixels) against the input list of artificial stars. The detection completeness was computed as a function of the original, input magnitudes of the artificial sources as the fraction of artificial sources within an input magnitude range [m1,m2] that were recovered by SExtractor (Oesch et al. 2007). Figure 14 reports the results of this analysis, which has been performed over the whole area covered by the survey.

Contamination by false detection has been examined using the negative image, with the same SExtractor configuration file used for the positive image. We estimate a contamination rate of $\approx 7$ % in the positive image at magnitude $U_V = 29.75$ (see Figure 15).

The $R_V$ band mosaic consists of images collected by different programs and has significantly non-uniform exposure time (see Figure 2). The depth of the image is therefore also quite non-uniform. Within the footprint of the ACS GOODS data, a fluctuation analysis like that described above gives an average $1\sigma$ magnitude limit $R_{lim} \approx 29.25$ AB mag within $1''$ radius apertures. The average detection completeness within the same area is illustrated in Figure 16.

### 3. Photometric Selection of Lyman Break Galaxies at z ≈ 3

Lyman Break Galaxies have unique colors due to the Lyman break at 912Å and Lyman $\alpha$ absorption blueward of 1216Å. In GOODS, until now, the detailed analysis of $z \approx 3$ galaxies has been limited to the bright part of the luminosity function due to the lack of deep coverage in the $U$ band. The best U band datasets available for GOODS until now come from KPNO 4m MOSAIC observations for GOODS-North (Capak et al. 2004), and from CTIO 4m MOSAIC (Dahlen et al. 2007) and ESO WFI observations (Hildebrandt et al. 2005, 2007) for GOODS-South.

Here we use the $U_V$, $B_{435}$ (Giavalisco et al. 2004), and $R_V$ images to perform a preliminary selection of LBG, and we compare the results with the available FORS2 and VIMOS spectroscopy. In the area covered by ACS, these data sets have $5\sigma$ depths of $\approx 28.0$ AB in the $U_V$ band ($1''$ radius aperture), 26.8 AB in $B_{435}$ ($0.5''$ radius), and 27.5 AB in R ($1''$ radius), including the effects of noise correlation.

In order to carry out matched-aperture photometry, we drizzled the $B_{435}$ image tiles[19] onto the same astrometric grid defined by the $R_V$ and $U_V$ mosaic images. We used SExtractor to detect galaxies in the $R_V$ band mosaic, and measured photometry through matched apertures in the $B_{435}$ and $U_V$ data. The colors of detected objects were computed using an aperture radius of $1''$ for the $R_V$ and $U_V$ images, and an aperture radius of $0.5''$ for the $B_{435}$ data. To account for the different PSFs in the three images, we determined an aperture correction to a total magnitude within a $5''$ radius, using measurements of bright point sources in each band. We obtained aperture corrections of 0.18, 0.1, and 0.13 magnitudes for the $U_V$, $B_{435}$ and $R_V$ images, respectively, which we then applied to the photometry for all sources.

The filter set used here is different from that used by Steidel et al. (2003) for their LBG color selection (see Figure 17). The $U_V$ band is narrower and redder than the Steidel $U_n$ filter. To define the locus in the $(B_{435}$-$R_V)$ vs. $(U_V$-$B_{435})$ color-color plane where star–forming galaxies at $z \sim 3$ are found, maximizing the inclusion of intrinsic UV colors while minimizing contamination from foreground interlopers, we have compared our photometric catalogs to existing spectroscopic data in the GOODS-South field, and have also carried out simulations using artificial objects to determine the color selection efficiency. We describe each of these in turn here.

We matched our VIMOS + ACS photometric catalogs against redshift lists derived from several different spectroscopic campaigns. The results are shown in Figure 18, where only galaxies with $23.5 \leq R_{MAG\_AUTO} \leq 27.0$ have been plotted. Photometric error bars for a color $C \equiv b1 - b2$ (where $b1$ and $b2$ are magnitudes in two bands) are computed as $\sigma_c = \sqrt{\sigma_{b1}^2 + \sigma_{b2}^2}$. Our adopted LBG color selection box is defined by

$$U_V - B_{435} \geq 0.56 \times (B_{435} - R) + 0.21, \quad 0.35 \leq B_{435} - R \leq 2.15 \tag{4}$$

$$U_V - B_{435} \geq 2.30 \times (B_{435} - R) - 3.54, \quad 2.15 \leq B_{435} - R \leq 3.50 \tag{5}$$

---

[19]http://archive.stsci.edu/prepds/goods/

with the further conditions $\sigma(R_{MAG\_AUTO}) \leq 0.1$ and $\sigma(B_{435,MAG\_APER}) \leq 0.5$. The green circles indicate LBG galaxies observed during the VIMOS spectroscopic campaign (Popesso et al. 2009), which were selected using CTIO U,WFI B and WFI R datasets, with spectroscopic redshift $\geq 2.8$, and galaxies from the FORS2 GOODS spectroscopic campaign with $2.8 \leq z \leq 3.5$. Yellow circles indicate galaxies with VIMOS spectroscopic redshift $2.2 \leq z \leq 2.8$, while red circle are objects with spectroscopic redshifts in the range $1.8 \leq z \leq 2.2$. Blue squares show foreground galaxies at $z \leq 1.8$ from the FORS2 GOODS, (Vanzella et al. 2008), K20, (Cimatti et al. 2002) spectroscopic campaigns, and from the VIMOS Medium Resolution GOODS spectroscopic campaign (Popesso et al. 2009). The selection box is thus quite efficient for selecting galaxies with $z \geq 2.8$: it includes one interloper, a $z = 0.9$ galaxy.

There are 1179 objects in the catalog which satisfy the LBG selection criteria (Figure 19), where objects with $S/N < 1$ in the $U_V$ band were assigned a magnitude limit $U_{lim} = 30$, approximately the $1\sigma$ sensitivity of the data, when calculating their $U_V - B_{435}$ colors. This gives a surface density of 7.3 arcmin$^{-2}$ over the $B_{435}$ area. Note that this surface density is purely indicative due to the spatially variable depth of the $R_V$ data. In Figure 20 we report the LBG number counts, compared with measurements by Capak et al. (2004) and Steidel et al. (1999). The plot is again indicative due to the different U filters used, with $U_V$ having the reddest cutoff.

We have used artificial object simulations to determine the completeness of the LBG color selection and to estimate the redshift distribution of the selected U dropout samples. We generate synthetic colors for model LBG as a function of redshift and for various stellar population parameters using the evolutionary population synthesis models of Bruzual & Charlot (2003, henceforth BC03). The model galaxy spectra were generated assuming constant star formation rates with ages from 10 to 300 Myr, and were reddened with the Calzetti et al. (2000) dust attenuation law, with E(B-V) in the range 0.0 to 0.5 mag. Various studies (e.g., Reddy et al. 2008) have shown that these parameters reliably reproduce the range of UV rest-frame colors observed for Lyman break galaxies. The models use the evolutionary tracks dubbed Padova 1994, with solar metallicity (Shapley et al. 2004), and a Salpeter IMF. There is very little difference in the simulated colors if a Chabrier IMF is used instead, particularly in the UV rest frame which is dominated by blue, high-mass stars. Each simulated source is assigned a random redshift between 2.0 and 4.5. The opacity of the intergalactic medium was calculated using recipe by Madau (1995). Meiksin (2006) has proposed larger attenuation than that of the Madau prescription, but we have used the Madau model in order to facilitate comparison with other surveys. Once obscured by the effects of dust, redshifted and attenuated by the cosmic opacity, the SED models have been multiplied by the system response function of the atmosphere, telescope and instrument (including detector and passband) to finally compute the observed broad–band colors.

To estimate the redshift selection efficiency, we inserted artificial objects into the $R_V$, ACS B, and $U_V$ band images, limiting the simulation to regions with effective $R_V$ exposure time $\geq 20 ks$. The colors and magnitudes of the artificial galaxies were drawn from the LBG simulations described above. We then run SExtractor and match the detected sources against the input list of artificial objects. In each redshift interval $(z_1, z_2)$ and magnitude interval $(m_1, m_2)$, we consider the simulated LBG whose true (noiseless, input) colors are within our LBG selection window, and calculate the fraction of these which are recovered in the SExtractor catalog with colors still falling within the LBG color box, and with $\sigma(R_{MAG\_AUTO}) \leq 0.1$ and $\sigma(B_{MAG\_APER}) \leq 0.5$. Note that this definition of completeness does not consider all possible galaxies at a given redshift, but only those whose intrinsic UV colors are typical of LBG as determined from previous surveys.

Figure 21 shows the redshift distribution of the recovered LBG, integrated over the magnitude range $23.5 \leq R_{AB,MAG\_AUTO} \leq 27.5$. This should be taken as indicative only, since no attempt was made to introduce a realistic luminosity function and hence apparent magnitude distribution for the simulated LBG. However, as expected due to the redder $U_V$ filter, the redshift distribution is skewed toward somewhat higher redshifts than is the case for the UGR color selection of Steidel et al. The color criteria primarily select galaxies at $z > 2.9$, with a peak at $z \approx 3.25$, and a significant tail extending to $z \approx 4$. Figure 22 shows the LBG selection completeness versus $R_V$ magnitude, considering only the redshift range $3.0 < z < 3.6$, where the redshift distribution peaks. The LBG selection is >70% complete down to nearly $R = 25.5$. The completeness falls off at fainter magnitudes, but is still >40% at $R = 26.0$ and 20% at $R = 26.5$ (where the typical S/N in the $R_V$ band is $\sim 10$, i.e., roughly the limit defined by our criterion $\sigma(R_{MAG\_AUTO}) \leq 0.1$, specified above). This is substantially fainter than the U dropout color selection of Steidel et al. (1999) using shallower data, which was highly incomplete by $R = 25.5$.

We analyze the effects of photometric errors in the definition of the LBG selection box. We fit a polynomial to the aperture error for all objects from the catalog with $23.5 \leq R_{MAG\_AUTO} \leq 27.0$ and $\sigma(R_{MAG\_AUTO}) \leq 0.1$. The results are reported in Figure 23 (only 1 out of 5 objects plotted). We then added errors to the simulations used for the definition of the selection box. For each artificial magnitude, in a given band, the error has been drawn from a normal distribution with a mean given by the fit at that magnitude and standard deviation dependent on the magnitude. Figure 23 shows that $z \approx 2.8$ galaxies tend to leave the selection box, while of course lower redshift objects tend to enter it.

## 4. Photometric redshifts

We have tested the added value of the $U_V$ band data for photometric redshift estimation in the GOODS South field. We have used the publicly available GOODS-MUSIC catalog (Grazian et al. 2006), with the improved photometry for the IRAC bands (see Santini et al. 2009 for a detailed description of the version 2 of this catalog) We have thus collected a sample of 1053 galaxies with good quality spectroscopic redshifts in the GOODS-South area covered by $U_V$ imaging with exposure time grater than 50 ks.

The photometric redshifts for all galaxies in the GOODS-MUSIC V2 catalog have been computed through a standard $\chi^2$ minimization procedure applied to the observed galaxy SED, and using photometric data from the VIMOS $U_V$, ACS $B$, $V$, $i$, and $z$, VLT-ISAAC $J$, $J$, and $K_s$, and IRAC $3.6\mu,4.5\mu,5.8\mu$, and $8\mu$ bands. The adopted spectral library of galaxies is based on the synthetic templates of PEGASE 2.0 (Fioc & Rocca–Volmerange 1997). Details about our photometric redshift calculations can be found in Giallongo et al. (1998), Fontana et al. (2000), Fontana et al. (2003), Fontana et al. (2004) and in Grazian et al. (2006).

Figure 25 shows the comparison between the spectroscopic and photometric redshifts derived with the $U_V$ photometry. The scatter of $z_{spec} - z_{phot}$ is $\sigma_z = 0.085$, with 22 outliers (defined as galaxies with $|z_{spec} - z_{phot}| \geq 0.5$). In the inset the histogram of $\frac{\Delta z}{1+z}$ is shown. In Figure 26, photometric redshifts have been derived without $U_V$ photometry. This results in an increase of the scatter, ($\sigma_z = 0.094$) and the number of outliers is significantly enhanced (41%).

The improvement is particularly relevant in the redshift range $2 \leq z \leq 4$ where the inclusion of the $U_V$ band photometry gives a mean $|z_{spec} - z_{phot}|$ of -0.002 ($\sigma_z = 0.173$) versus a mean of -0.110 ($\sigma_z = 0.167$) obtained with the exclusion of $U_V$ band data.

## 5. Summary and Conclusions

We have presented deep $U_V$ band imaging data of the GOODS-South field, collected in the framework of the ESO GOODS Public Survey. Here we summarize the main results.

• We present a photometrically and astrometrically calibrated stack of $U_V$ band data over GOODS-South, covering an area of $\approx 625$ arcmin$^2$. The depth and overall image quality of the final coadded $U_V$ data, $\approx 30AB$ at $1\sigma$ over the ACS area, match the already impressive multiwavelength data coverage of GOODS-South.

• In order to facilitate selection of LBG at $z \approx 3$, we also present a moderately deep

photometrically and astrometrically calibrated stack of $R_V$ band data, which, in its deepest region covering $\approx 90$ arcmin$^2$, is capable of detecting $z \approx 3$ LBG as faint as $M_{AB}(1700\text{Å}) = -18.5$. There have been other multi-color data sets for $U$–dropout LBG selection that reach somewhat deeper photometric limits (Sawicki & Thompson 2006), or similar depth over wider areas (e.g., the CFHT Legacy Survey, Hildebrandt et al. 2009). However, there is unique value to having such data available in the GOODS fields, where the depth and breadth of the other multiwavelength data (e.g., from HST, Chandra and Spitzer) are unique and provide extensive opportunities to investigate various astrophysical properties of faint LBG. We note that careful comparison between multicolor datasets for LBG selection must take into account the different filters used, in particular the U band filter. The long wavelength cutoff of the VIMOS $U_V$ filter is $\approx 250\text{Å}$ redder than that of the $U_n$ used by Steidel et al. (2003) and Sawicki & Thompson (2006), while the cutoff of the CFHTLS $u^*$ filter is $\approx 200\text{Å}$ redder than the VIMOS $U_V$ (see Figure 7) .

- The $U_V$ band data presented here are also valuable for reducing uncertainties in photometric redshifts. Using the public available GOODS-MUSIC catalog (Grazian et al. 2006), we demonstrate that the $U_V$ band data improve the accuracy of photometric redshifts and reduce the catastrophic error rate, especially in the redshift interval $2 \leq z \leq 4$.

## Acknowledgments

We thank the Paranal staff for the effort in collecting the images presented here. We thank Ricardo Demarco for the help in retrieving SA98 spectrophotometric data. We acknowledge the financial contribution from contract ASI I/016/07/O and from the PRIN INAF "*A deep VLT and LBT view of the Early Universe: the physics of high-redshift galaxies.*". This work was partially supported by the ESO Director General Discretionary Funds

## Appendix A. Astrometric solution

We summarize the method used to derive the astrometric solution and its mapping to the SIP convention (Shupe et al. 2005). The plate model (e.g., Platais et al. 2002) used here is:

$$\xi = \sum_{i,j=0}^{N} a_{ij} x^i y^j$$

$$\eta = \sum_{i,j=0}^{N} b_{ij} x^i y^j$$

where $\xi$ and $\eta$ are the standard coordinates derived from the gnomonic projection and x and y are $x_{CCD} - CRPIX1$ and $y_{CCD} - CRPIX2$ respectively. The coefficients $a_{ij}$ and $b_{ij}$ (which are different for the four chips) have been derived using the matching of the extracted sources (via triangulation, e.g., Valdes et al. 1995) with the external reference catalog, *and* the matching among the sources from images from the same set of consecutive and dithered exposures within a single Observation Block (OB,Chavan et al. 1998) (and on the same chip). The constraint of uninterrupted OB comes from the assumption that the $a$ and $b$ coefficients are constant within the OB. With the further assumption that the distortion polynomial is of second order, the solution of the derived *overdetermined* system is obtained minimizing the $\chi^2$ given by the sum of the terms:

$$\chi^2_{\alpha,ref} = \sum_{c=1}^{N} \sum_r \frac{\|\alpha(x_{c,r}y_{c,r}) - \alpha(r)\|^2}{\sigma^2_{cat} + \sigma^2_{obs,c}}$$

$$\chi^2_{\delta,ref} = \sum_{c=1}^{N} \sum_r \frac{\|\delta(x_{c,r}y_{c,r}) - \delta(r)\|^2}{\sigma^2_{cat} + \sigma^2_{obs,c}}$$

$$\chi^2_{\alpha,overlap} = \sum_{i=1}^{N-1} \sum_{j=i+1}^{N} \sum_o \frac{\|\alpha(x_{i,o}y_{i,o}) - \alpha(x_{j,o}y_{j,o})\|^2}{\sigma^2_{i,o} + \sigma^2_{j,o}}$$

$$\chi^2_{\delta,overlap} = \sum_{i=1}^{N} \sum_{j=i+1}^{N} \sum_o \frac{\|\delta(x_{i,o}y_{i,o}) - \delta(x_{j,o}y_{j,o})\|^2}{\sigma^2_{i,o} + \sigma^2_{j,o}}$$

where $\alpha(r), \delta(r), \alpha(x,y), \delta(x,y)$ refer to the center of the field. $\sigma_{cat}$ and $\sigma_{obs,c}, \sigma_{i,o}, \sigma_{j,o}$ are the errors in the position in the reference catalog and in the extracted sources respectively, and N is the number of exposures in the OB. The minimization of the total $\chi^2$ corresponds to the minimization of $\|\mathbf{A} \times \mathbf{x} - \mathbf{y}\|^2$, which is done via Singular Value Decomposition. Here $\mathbf{A}$ is the design matrix, with dimensions given by *(number_of_reference+number_overlaps) × (number_of_coefficients)*, $\mathbf{x}$ the vector of the coefficients to be determined and $\mathbf{y}$ the vector of the observations. Since by construction the $\xi$ and $\eta$ distortions are independent, the $\chi^2$ minimization can be solved separately, in order to obtain the $(a_{00i}, a_{10}, a_{01}, a_{11}, a_{20}, a_{02})$ and $(b_{00i}, b_{10}, b_{01}, b_{11}, b_{20}, b_{02})$ coefficients. The $a_{00i}$ and $b_{00i}$ give the correction for the CRVAL1 and CRVAL2 respectively for exposure $i$. The $a_{10}, a_{01}, b_{10}, b_{01}$ terms correspond to the more common CD1_1,CD1_2,CD2_1,CD2_2 coefficients.

This SIP convention is such that

$$\begin{pmatrix} \alpha \\ \delta \end{pmatrix} = \begin{pmatrix} CD1\_1 & CD1\_2 \\ CD2\_1 & CD2\_2 \end{pmatrix} \begin{pmatrix} x + f(x,y) \\ y + g(x,y) \end{pmatrix}$$

where $x = X_{CCD} - CRPIX1$ and $y = Y_{CCD} - CRPIX2$ with $X_{CCD}$ and $Y_{CCD}$ begin the measured x and y centroid position on the chip. $f(x,y)$ and $g(x,y)$ are the quadratic and higher order terms of the distortion polynomial:

$$f(x,y) = \sum_{i,j} A\_i\_j x^i y^j \qquad 2 \leq i+j \leq A\_ORDER$$

$$g(x,y) = \sum_{i,j} B\_i\_j x^i y^j \qquad 2 \leq i+j \leq B\_ORDER$$

A_ORDER and B_ORDER are the polynomial distortion order in x and y respectively. In the astrometric solution for the $U_V$ and $R_V$ data bands a second order polynomial has been used: $f(x, y) = A\_2\_0x^2 + A\_1\_1xy + A\_0\_2y^2$

$g(x, y) = B\_2\_0x^2 + B\_1\_1xy + B\_0\_2y^2$

From the vectors $a$ and $b$ found previously the $A\_i\_j$ and $B\_i\_j$ terms can be easily found (e.g., $A\_2\_0 = \frac{a_{20}CD2\_2 - b_{20}CD1\_2}{CD1\_1CD2\_2 - CD1\_2CD2\_1}$) and these values are directly inserted into the image header as value to the corresponding keywords. This formulation of the distortion polynomials also allows us to estimate the change in the pixel area from center to corners, which amounts to $\leq 2\%$.

## Appendix B. Relative Photometry

The photometric scaling for the single frames has been obtained as the product of the relative photometry within a tile (in the R band only for the data from the 167.D-0492 programme) and the relative photometry among the different tiles. The inter-tile relative photometry has been obtained following Koranyi et al. (1998). Given a set of $N$ tiles, the scaling terms are given by the minimization of

$$\chi^2 = \sum_{i=1}^{N-1} \sum_{j=i+1}^{N} \left\| \frac{(zpt_i - zpt_j - \delta_{ij})}{\sigma_{ij}} \right\|^2$$

where $\delta_{ij}$ is the median difference among the matched sources between tile $i$ and tile $j$ (if any), and $\sigma_{ij}$ the corresponding standard deviation. Minimizing this $\chi^2$ is equivalent to the minimization of $\mathbf{A} \times \mathbf{z} = \mathbf{y}$, where the design matrix $\mathbf{A}$ is such that $a_{ij} = 0$ if there is no overlap between tile $i$ and tile $j$, and $a_{ij} = -a_{ji} = \sigma_{ij}^{-2}$ if there is overlap, $z$ the vector of zero points and $y$ the vector of the N observed differences. $\mathbf{A}$ is thus a sparse antisymmetric matrix and the solution for $z$ can be found using e.g., $ATLAS$ [20]. Since VIMOS is a 4 chip camera with each chip having its own filter, a basic assumption for this inter-tile relative photometric calibration step is that both the color term *and* the extinction color term are constant (for the given band) in time and that they have the same value for the four chips. To avoid singularities, for one of the tiles the $zpt$ has been set to 0.

## Appendix C. VIMOS $U_V$ filter red leak

The $U_V$ filter in VIMOS quadrant 4, named *vm-U-4.2*, is known to have a red leak (see Figure 27). To examine the effect of this red leak on LBG selection, we repeat simulations similar to those used for the definition of the LBG color selection box, computing the color difference with and without the red leak, $\Delta U = U_V(\text{red leak}) - U_V(\text{no red leak})$. The results are plotted in Figure 24: the effect starts to be significant ($\geq 0.1 mag$) at redshift $\geq 4$. To test the possibility that the red leak transforms a B-dropout into an $U_V$-dropout, we generated BC03 models in a similar way to what was done for the $z \approx 3$ LBG, and select them according to B-dropout selection criteria of Bouwens et al. (2007). For each object we also computed $\Delta U_V = U_V(\text{red leak}) - U_V(\text{no red leak})$. Even if the effect is as strong at $z \geq 4$ as the simulations suggest, i.e., $\geq 0.5 mag$ in $\Delta U_V$, we find that this is not enough to move the B-dropout into the U-dropout color selection box.

---

[20]http://math-atlas.sourceforge.net/

### Appendix D. The undecimated wavelet transform

The development of the wavelet theory has made multi-scale methods very popular in image processing application, (Mallat 1989). Within this theory a key role in the decomposition algorithms is played by the *analysis scaling* function $\phi(x)$, which has to satisfy , along with other conditions, the refinement relation:

$$\phi(x) = 2 \sum_{k \in Z} h(k) \phi(2x - k)$$

where $h$ is the analysis filter. In this paper we used B-splines as scaling functions. (Unser & Blu 2003) has also shown that any admissible scaling function can be expressed as a convolution of a B-spline and a distribution. As a consequence B-splines are responsible of the smooth part behaviour of $\phi$, and thus they are a natural choice as a scaling function when searching for smooth components. In one dimension, a symmetrical B-spline of order $n$ can be expressed by

$$\beta^n(x) := \sum_{j=0}^{n+1} \frac{(-1)^j}{n!} \left[ \binom{m+n}{n} \right] \left( x + \frac{n+1}{2} - j \right)^n \times \theta\left( x + \frac{n+1}{2} - j \right) \ (x \in R)$$

where $\theta(x)$ is the Heaviside step function. For $n = 3$, one gets the cubic B-spline $\beta^3(x) = \frac{1}{6}((x+2)^3 \times \theta(x+2) - 4(+1)^3 \times \theta(x+1) + 6x^3 \times \theta(x) - 4(x-1)^3 \times \theta(x-1) + (x-2)^3 \times \theta(x-2)$.

B-splines give also the advantage that the analysis scaling function has an explicit expression. For a cubic B-spline $\phi(x) = \beta^3(x)$, and using the refinement relation, the values for the analysis filter $h$ can be found:

$$h(k) = \frac{[1,4,6,4,1]}{16} \text{ with } k = -2, ..., 2$$

Since images are bi-dimensional, the refinement relation has to be recast in 2D, (Starck et al. 2007):

$$\phi(x,y) = \phi(x) \times \phi(y) = 4 \sum_{k \in Z} \sum_{l \in Z} h(k) h(l) \phi(2x - k) \phi(2y - l)$$

This means that $h(k,l) = h(k)h(l)$ is a $5 \times 5$ matrix:

$$\frac{1}{256} \cdot \begin{bmatrix} 1 & 4 & 6 & 4 & 1 \\ 4 & 16 & 24 & 16 & 4 \\ 6 & 24 & 36 & 24 & 6 \\ 4 & 16 & 24 & 16 & 4 \\ 1 & 4 & 6 & 4 & 1 \end{bmatrix}$$

which also satisfies $\sum_k \sum_l h(k,l) = 1$ thus the flux is conserved, under convolution with $h(k,l)$. At variance with the standard wavelet transform, the undecimated decomposition doesn't reduce the number of coefficient in the wavelet transform (decimation). Using the previous analysis filter $h(k,l)$ one gets the isotropic undecimated wavelet decomposition (Starck et al. 2007). The final product of this decomposition (up to level $J$) of an image $N \times M$ is a 3D array of dimensions $J \times N \times M$. This is clear if the image pixel values $I_0(n,m)$ are considered as the result of scalar product $I_0(n,m) = \langle f(x,y), \phi(x-n)\phi(y-m)\rangle$ At level $j \leq J$ the coefficient $I_j(n,m)$ of the decomposition is given by

$$I_j(n,m) = \langle f(x,y), 2^{-j}\phi(2^{-j} \cdot x - n) \cdot 2^{-j}\phi(2^{-j} \cdot y - m)\rangle$$

which can be considered a projection along vectors which are obtained by dilation (index $j$) and translation (indices $k,l$ of the original $\phi(x,y)$. Using the refinement relation it can be shown that

$$I_j(n,m) = \sum_k \sum_l h(k,l)I_{j-1}(n + 2^jk, m + 2^jl) \ (1 \leq j \leq J).$$

The difference between two consecutive resolutions $\omega_{j+1}(n,m) = I_j(n,m) - I_{j+1}(n,m)$ gives the wavelet coefficients at scale $j + 1$. Thus the initial image can be decomposed in

$$I_0(n,m) = I_J(n,m) + \sum_{j=1}^{J} \omega_{j+1}(n,m)$$

where $I_J(n,m)$ is the final smooth component, which in the present paper is used as an estimation of the background.

Table 1.   Log of Observations

| Date | Field ID | Total exposure time (seconds) | DIMM seeing range (arcsec) | Airmass range |
|------|----------|-------------------------------|----------------------------|---------------|
| 11-Aug-2004 | 01 | 3000 | 0.81-0.87 | 1.17-1.29 |
| 12-Nov-2004 | 01 | 3000 | 0.60-0.70 | 1.06-1.12 |
| 13-Nov-2004 | 01 | 4000 | 0.59-0.74 | 1.08-1.25 |
| 17-Nov-2004 | 01 | 3000 | 0.58-0.91 | 1.12-1.21 |
| 06-Sep-2005 | 02 | 6000 | 0.98-1.44 | 1.00-1.06 |
| 07-Sep-2005 | 02 | 4000 | 0.68-0.84 | 1.07-1.21 |
| 08-Sep-2005 (*) | 02 | 3000 | 0.99-1.03 | 1.01-1.03 |
| 08-Oct-2005 | 02 | 2000 | 0.78 | 1.07-1.11 |
| 10-Oct-2005 | 06 | 2000 | 0.58-0.74 | 1.07-1.10 |
| 11-Oct-2005 | 06 | 7000 | 0.37-0.82 | 1.00-1.08 |
| 12-Oct-2005 (*) | 03 | 3000 | 0.69-0.80 | 1.01-1.03 |
| 12-Oct-2005 (*) | 05 | 3000 | 0.64-0.74 | 1.05-1.11 |
| 12-Oct-2005 (*) | 06 | 3000 | 0.80-1.10 | 1.00-1.01 |
| 27-Oct-2005 | 02 | 3000 | 1.45-1.50 | 1.00-1.01 |
| 27-Oct-2005 | 03 | 6000 | 1.17-1.42 | 1.02-1.18 |
| 29-Oct-2005 (*) | 06 | 3000 | 0.90-1.21 | 1.02-1.06 |
| 29-Oct-2005 (*) | 07 | 4000 | 1.21-1.32 | 1.09-1.24 |
| 30-Oct-2005 | 03 | 3000 | 0.90-1.19 | 1.12-1.21 |
| 31-Oct-2005 | 03 | 5000 | 0.61-0.90 | 1.07-1.15 |
| 01-Nov-2005 | 03 | 1000 | 0.72 | 1.20-1.26 |
| 02-Dec-2005 | 03 | 3000 | 0.50-0.61 | 1.00-1.01 |
| 02-Dec-2005 | 05 | 6000 | 0.46-0.78 | 1.02-1.16 |
| 04-Dec-2005 (*) | 05 | 6000 | 0.64-1.04 | 1.00-1.02 |
| 04-Dec-2005 (*) | 07 | 6000 | 0.78-0.99 | 1.04-1.23 |
| 26-Jan-2006 | 07 | 3000 | 0.55-0.64 | 1.25-1.41 |
| 19-Aug-2006 (*) | 01 | 3000 | 1.22-1.47 | 1.03-1.06 |
| 22-Sep-2006 | 07 | 6000 | 0.79-1.30 | 1.00-1.04 |
| 24-Sep-2006 | 07 | 3000 | 0.95-1.12 | 1.09-1.16 |
| 24-Sep-2006 | 04 | 2000 | 0.72-0.78 | 1.02-1.04 |
| 25-Sep-2006 | 04 | 2000 | 0.39-0.48 | 1.02-1.04 |

Table 1—Continued

| Date | Field ID | Total exposure time (seconds) | DIMM seeing range (arcsec) | Airmass range |
|------|----------|-------------------------------|----------------------------|---------------|
| 13-Oct-2006 | 08 | 3000 | 0.85-1.14 | 1.05-1.12 |
| 16-Oct-2006 | 04 | 6000 | 0.51-0.72 | 1.01-1.13 |
| 16-Oct-2006 | 08 | 9000 | 0.54-0.75 | 1.00-1.22 |
| 17-Oct-2006 | 04 | 3000 | 0.79-0.83 | 1.07-1.13 |
| 18-Oct-2006 | 08 | 3000 | 0.88 | 1.09-1.17 |
| 21-Oct-2006 | 04 | 3000 | 0.45-0.52 | 1.02-1.06 |
| 27-Oct-2006 | 04 | 3000 | 0.50-0.66 | 1.14-1.24 |

Table 2.   $R_V$ band observations

| Program | Total exposure time (seconds) |
|---------|-------------------------------|
| 071.A-3036 | 7800 |
| 072.A-0586 | 600 |
| 074.A-0280 | 1330 |
| 074.A-0303 | 1170 |
| 074.A-0509 | 2700 |
| 075.A-0481 | 600 |
| 078.A-0485 | 5300 |
| 078.B-0425 | 600 |
| 080.A-0566 | 4770 |
| 080.A-0411 | 2550 |
| 167.D-0492 | 25920 |
| 171.A-3045 | 5400 |

Table 3.   Color and extinction terms

| $a_1$ | $a_2$ | $a_{B-V}$ | |
|-------|-------|-----------|---|
| 0.139 | -0.023 | 0.188 | $(B-V) \leq 0.0$ |
| 0.139 | -0.023 | -0.42×(B-V) | $(B-V) \geq 0.0$ *and* $(B-V) \leq 0.6$ |
| 0.139 | -0.023 | -0.0465+0.0266×(B-V) | $(B-V) \geq 0.6$ |

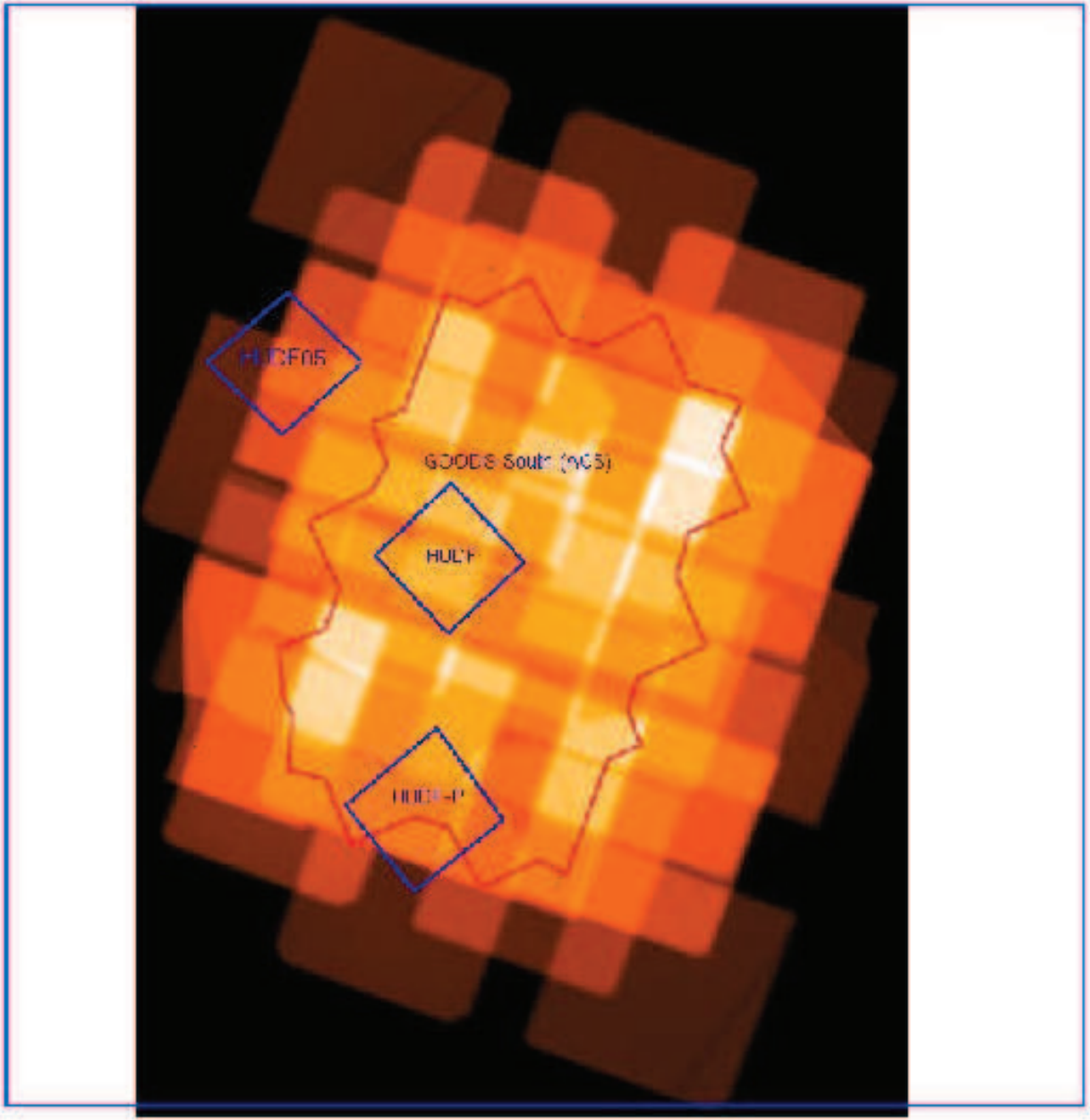

Fig. 1.— Exposure map of the $U_V$ band coadded image in the GOODS-South. Brighter areas correspond to deeper data. In the GOODS ACS region, the depth ranges from $U_V \approx 29.5$ to $U_V \approx 30.2$ mag (AB, $1\sigma$ in $1''$ radius). The footprints of the GOODS HST/ACS coverage and of HUDF, HUDF05 (Oesch et al. 2007), and of one of the HUDF parallel fields are shown. The outer box indicates the $\approx 30' \times 30'$ footprint of the Extended Chandra Deep Field South (ECDFS).

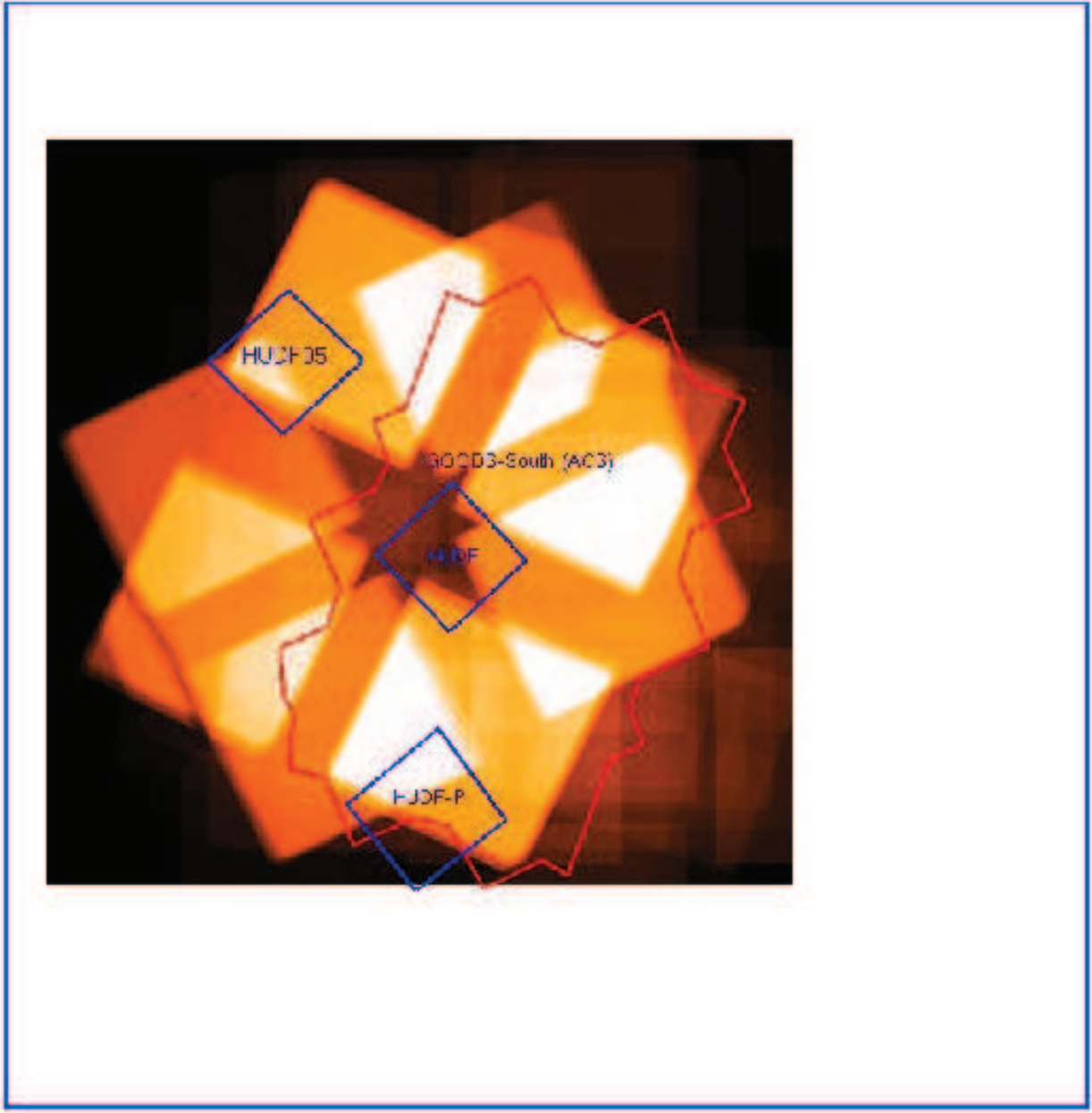

Fig. 2.— Exposure map of the $R_V$ band exposures in the GOODS-South. Brighter regions indicate deeper data. In the ACS area, the $R_V$ band depth ranges from 28.2 to 29.3 mag (AB, $1\sigma$ in $1''$ radius). The footprints of various HST imaging data sets are shown, as in Figure 1. The VIMOS $R_V$ coverage within the ACS area is less uniform than that of the $U_V$ band mosaic.

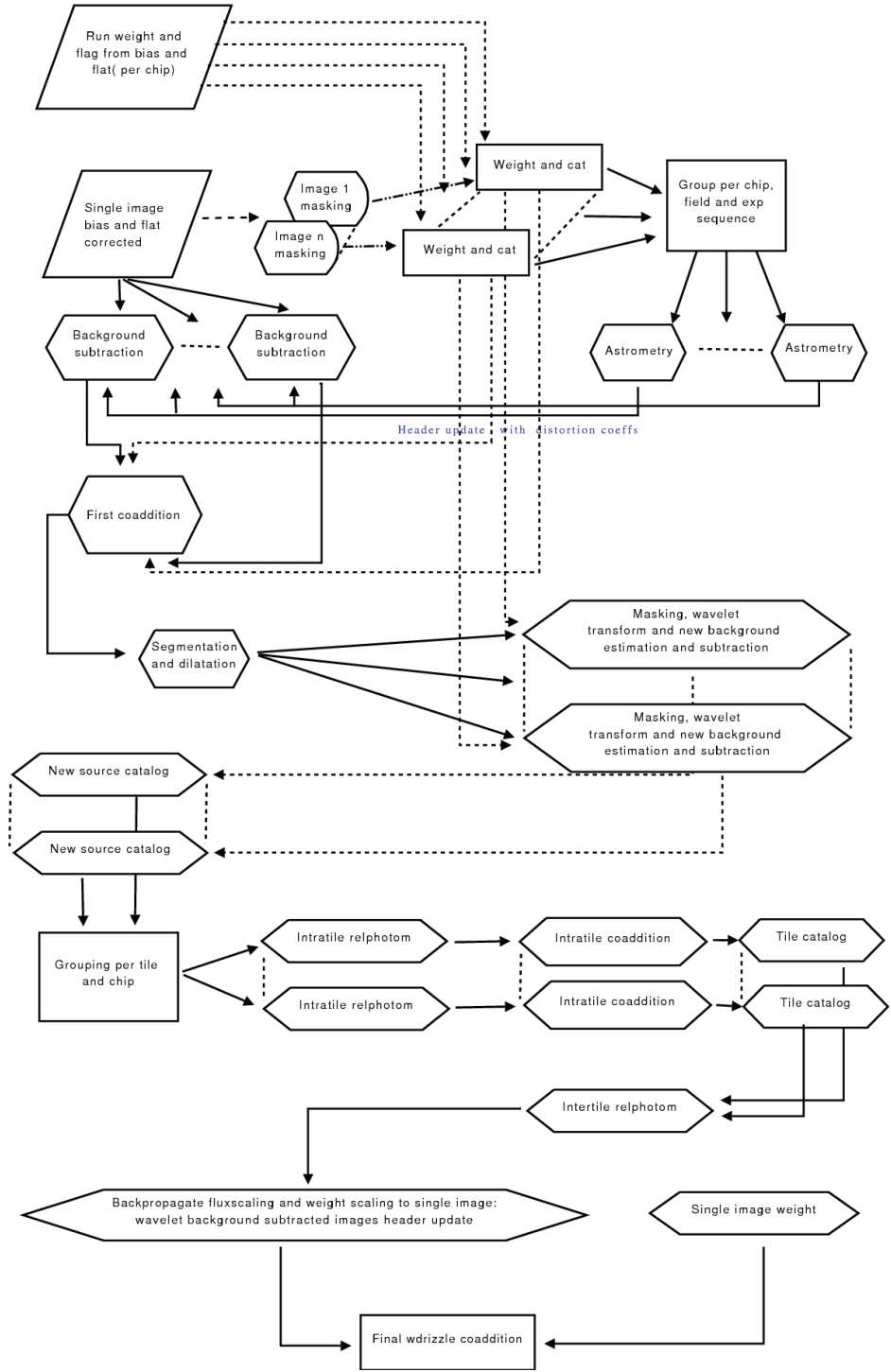

Fig. 3.— Flow Chart of the reduction steps followed for the $U_V$ band and $R_V$ band datasets.

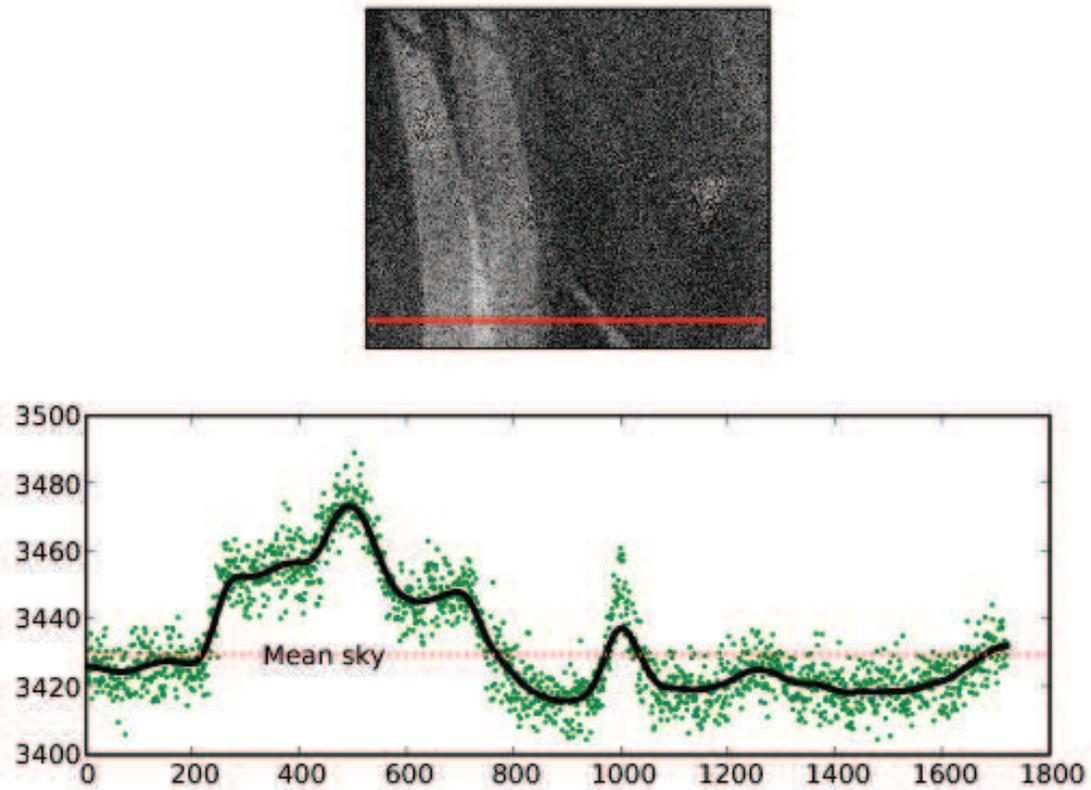

Fig. 4.— *Top:* Portion of a combined stack of 12 unregistered R band images, illustrating the typical diffuse light encountered in the data reduction. The red line indicates the cross section used for the bottom plot. The size of the region shown here is $\approx 5.5' \times 4.5'$. *Bottom:*. The points are the sky values in the cross section, while the thick line is the wavelet background estimation. The horizontal line is the global mean of the sky in the image shown.

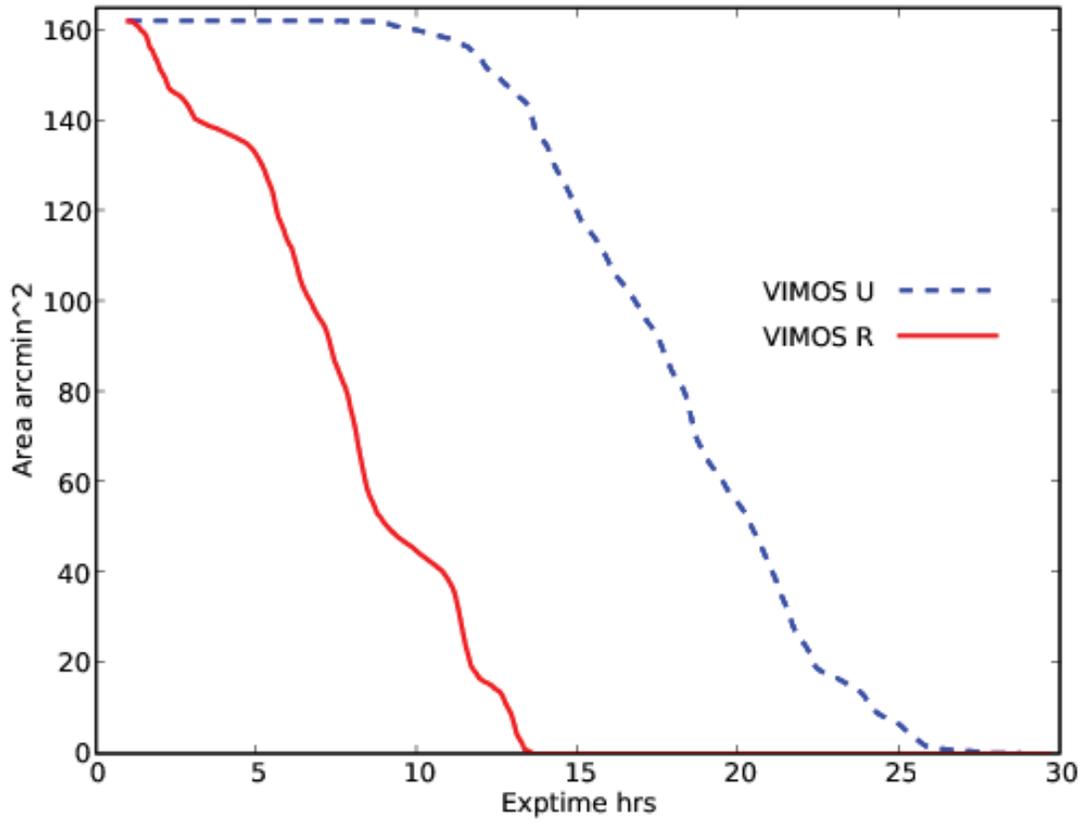

Fig. 5.— Cumulative distribution of the area covered as a function of the exposure time for the region also covered by the ACS-GOODS data.

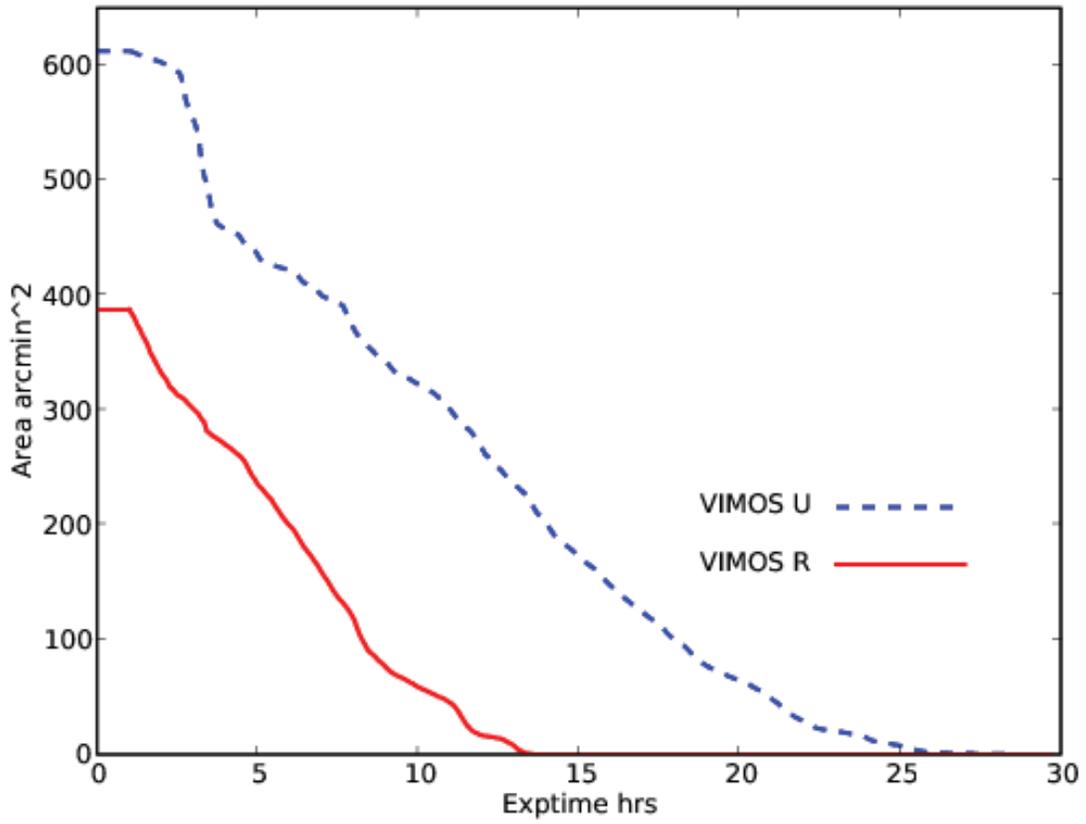

Fig. 6.— Cumulative distribution of the area covered as a function of the exposure time over the whole $U_V$ and $R_V$ mosaics.

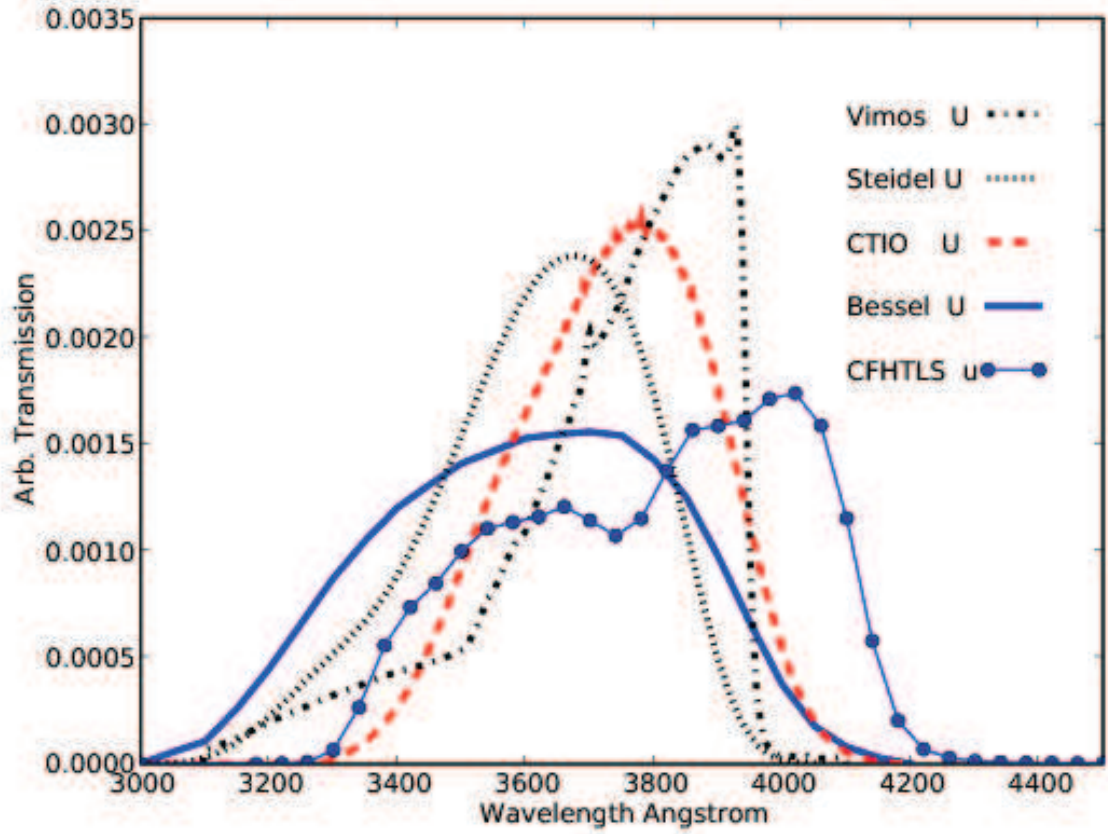

Fig. 7.— Comparison of the transmission (arbitrary units) among different U band filters: $U_V$ , Steidel $U_n$, CTIO U, CFHTLS u and Bessel U.

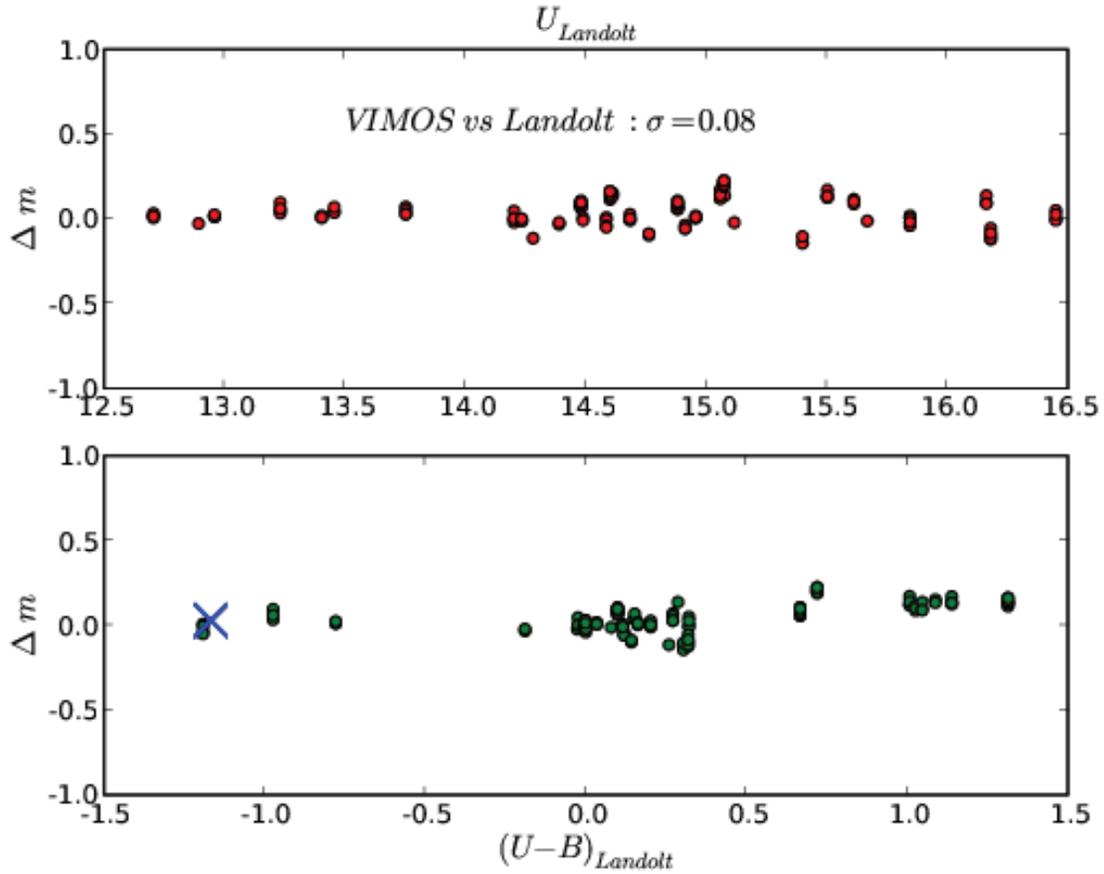

Fig. 8.— *Top panel*: Magnitude difference between Landolt (2007) U magnitude and the derived U magnitude from $U_V$ standard stars observations and reduction as a function of the literature magnitude. *Bottom panel:* The difference in magnitude is plotted against the U-B color of the standard stars. The blue cross indicates the spectrophotometric standard star Feige 110.

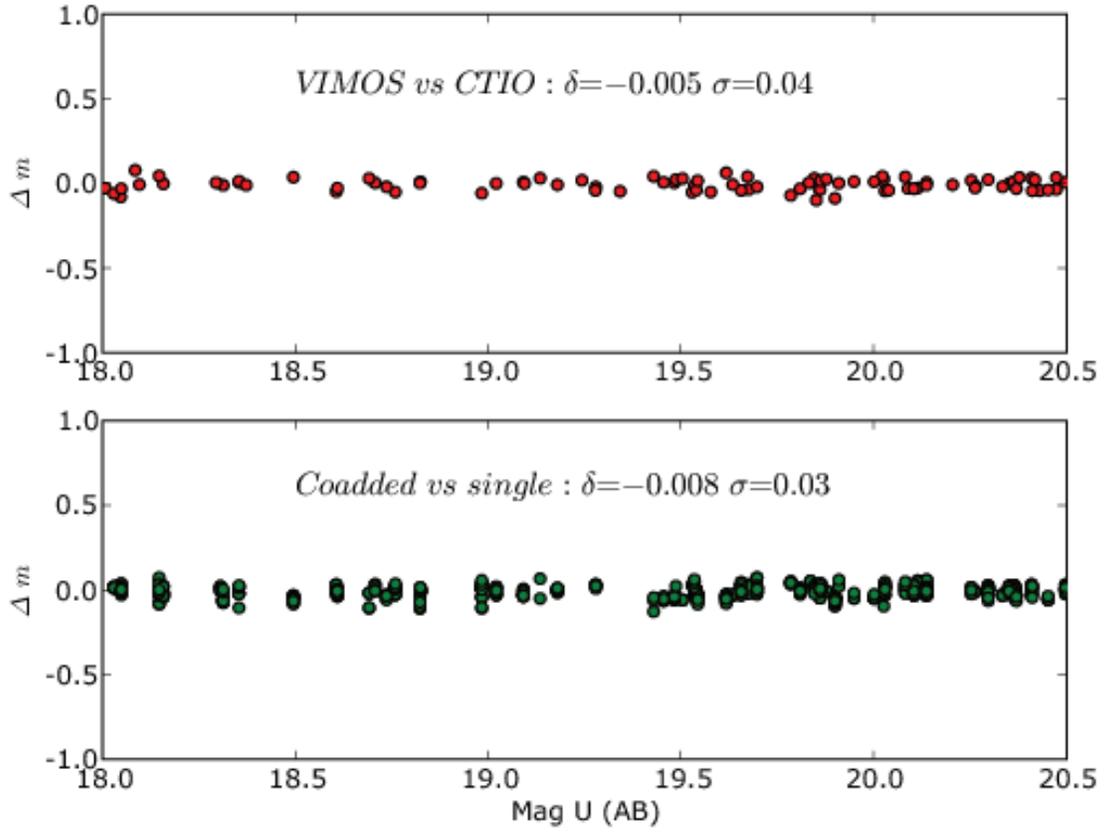

Fig. 9.— *Top panel:* aperture-corrected magnitude comparison between $U_V$ from the coadded image and the matched sources in CTIO U image. *Bottom panel:* Magnitude comparison of the same sources in the coadded $U_V$ mosaic versus an individual U band exposure.

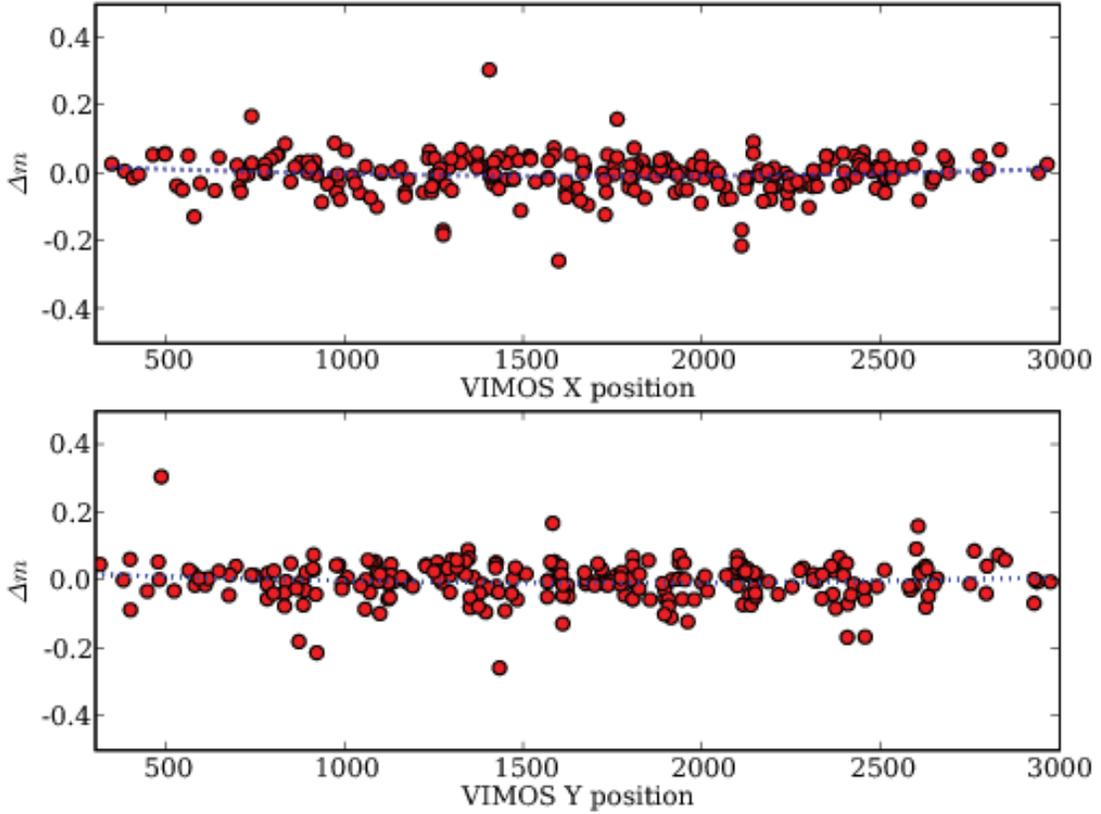

Fig. 10.— Comparison of $U_V$ band magnitudes from the different tiles and the CTIO U band magnitude for matched sources as a function of the x position (*top panel*) and y position (*bottom panel*) in the $U_V$ tiles. The dotted curves indicate the second order polynomial fitting to the differences.

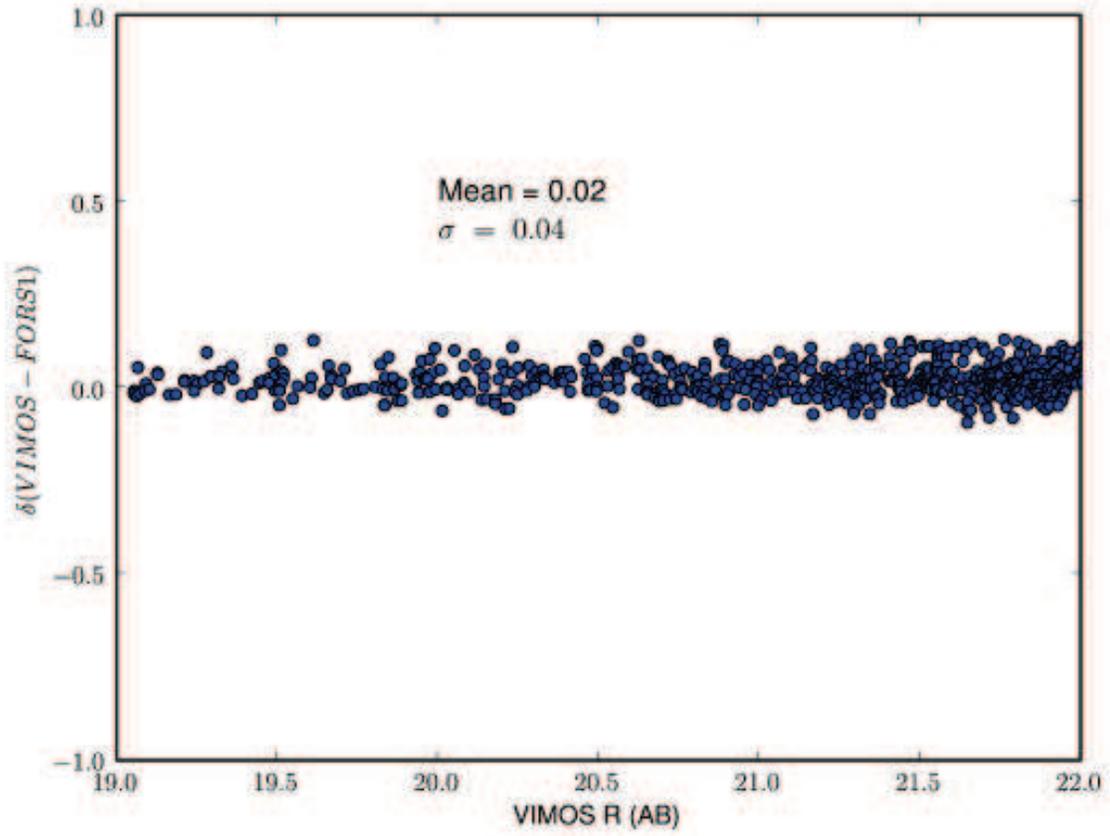

Fig. 11.— Magnitude differences for 650 objects in common between FORS1 R and $R_V$ mosaic.

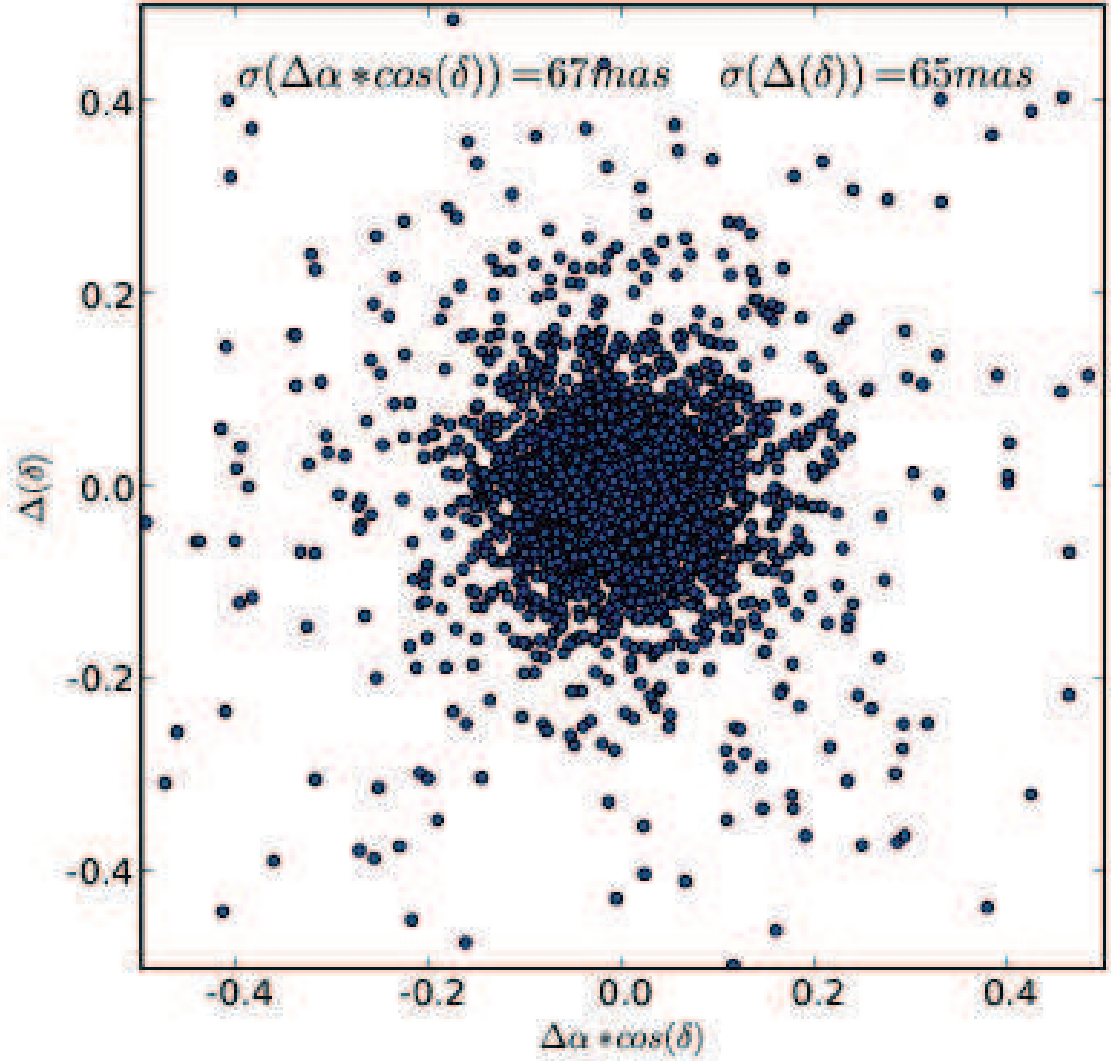

Fig. 12.— Difference in Right Ascension and Declination between $U_V$ band detected sources and the matched sources in the reference catalog from WFI R

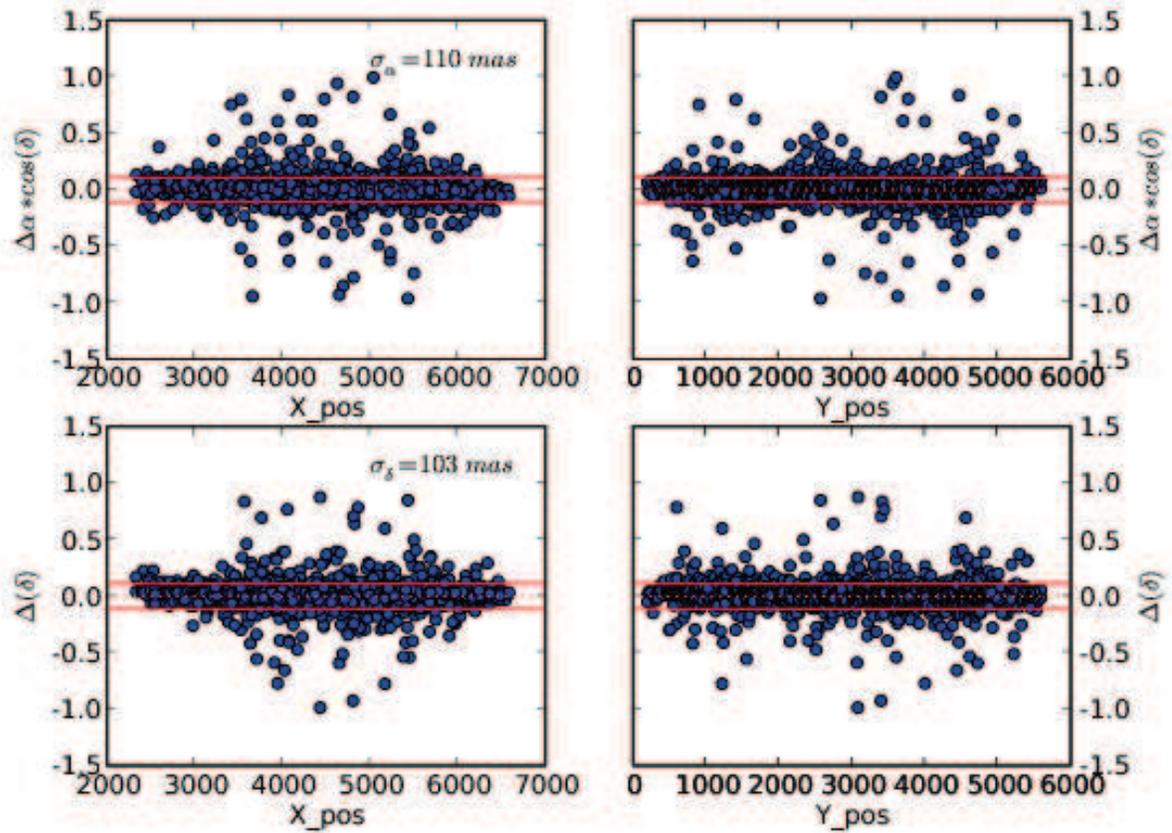

Fig. 13.— Plots of the difference in astrometry between U band and R band data obtained from matching independently extracted catalogs. The quoted $1\sigma$ dispersions have been derived from the data plotted without $\sigma$-clipping.

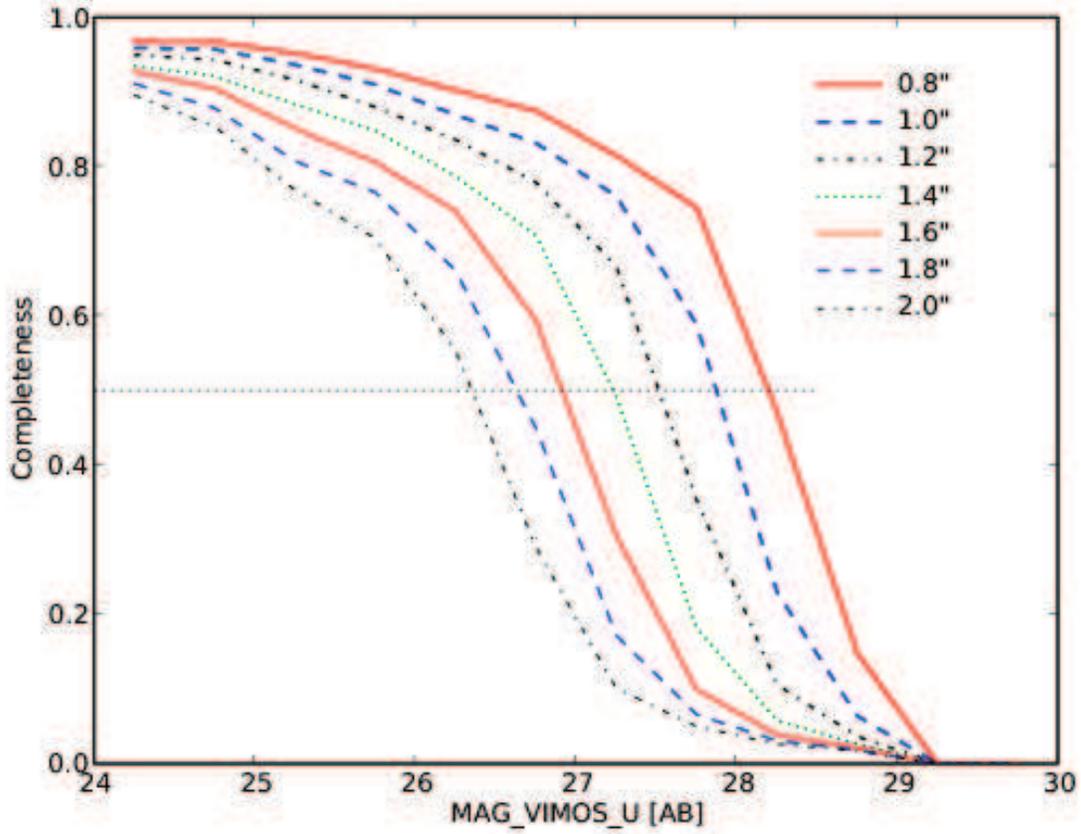

Fig. 14.— Completeness from simulations for the $U_V$ band coadded image over the area covered by the ACS $B_{435}$ data. Lines show the fraction of recovered over the implanted sources. The lines indicate results for simulated objects with different FWHM, ranging from 0.8 to 2.0 arcsec in steps of 0.2 arcsec, from right to left.

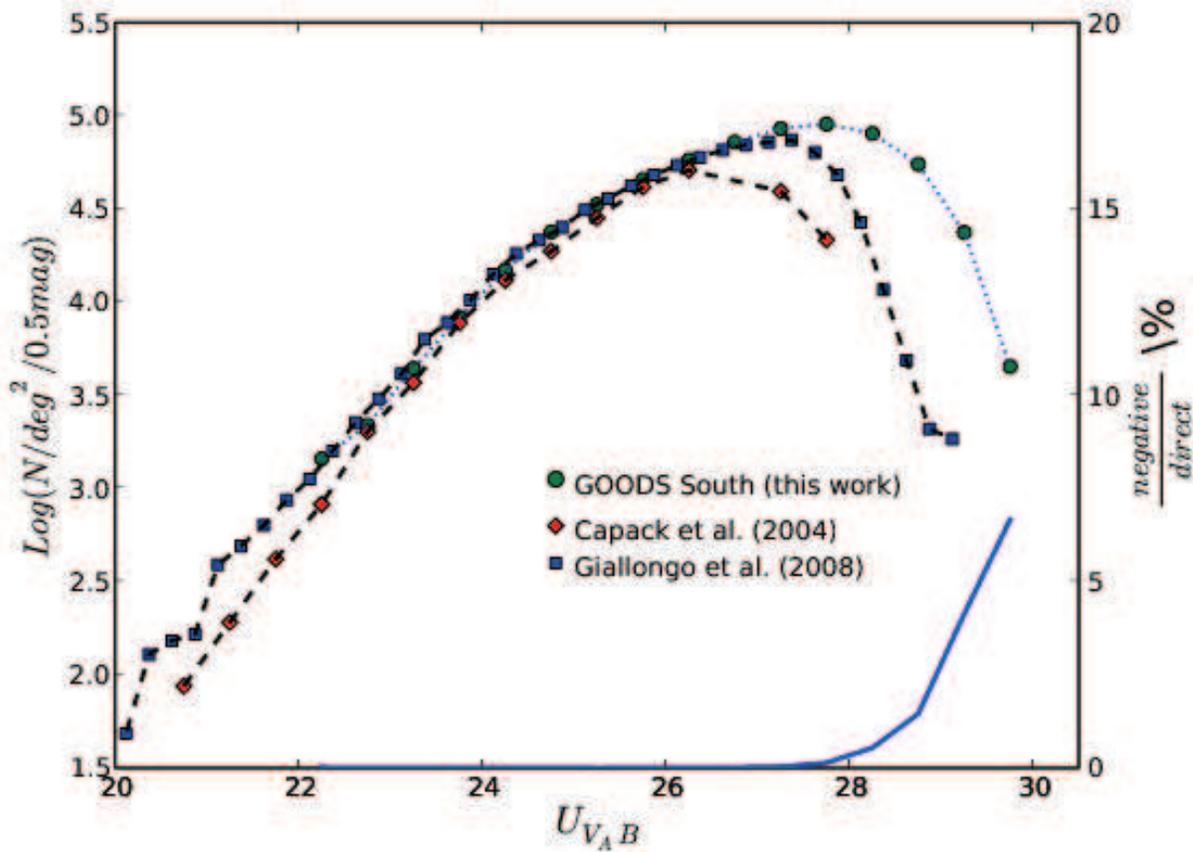

Fig. 15.— Comparison of number counts in the $U_V$ band for the GOODS-South, GOODS-North KPNO 4m MOSAIC data (Capak et al. 2004), and Large Binocular Telescope LBC data for the Q0933+28 field (Giallongo et al. 2008). All points are from raw, uncorrected data. The bottom curve shows the estimated fraction of spurious detections, computed from the ratio between the counts of "negative sources" in the inverted stacked image and those from the regular (positive) stacked image itself.

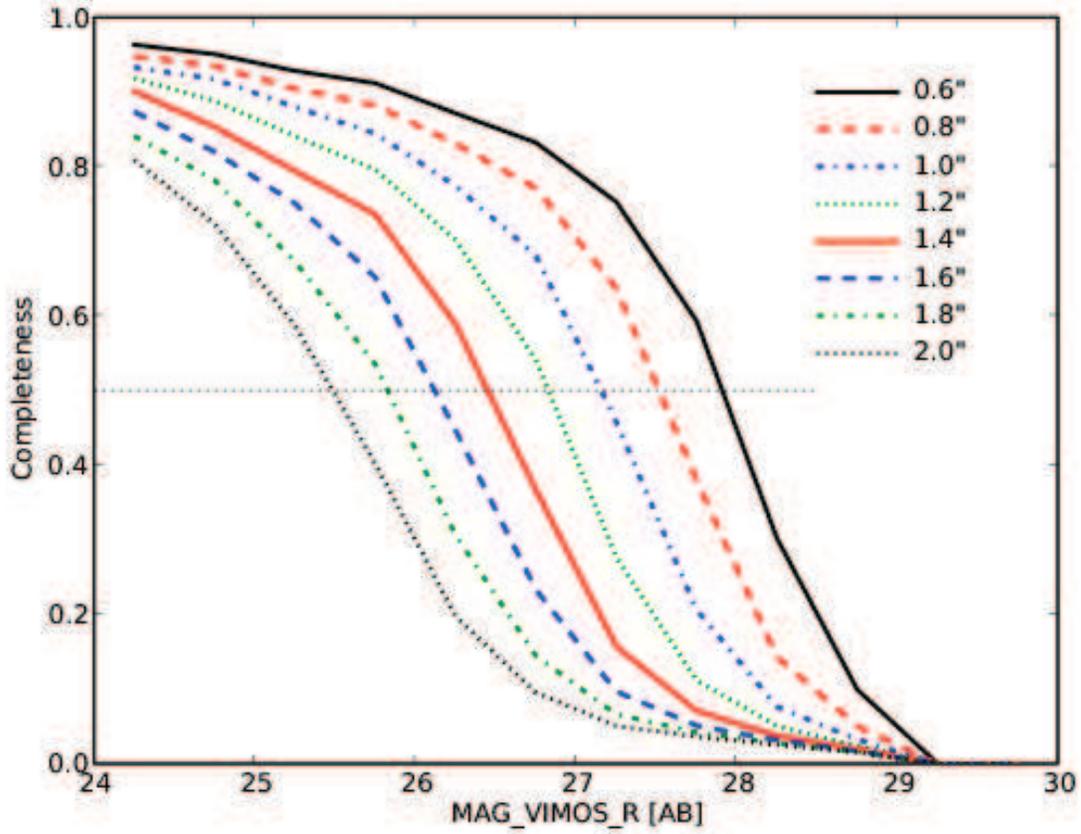

Fig. 16.— Completeness from simulations for the $R_V$ band coadded image over the area covered by the ACS $B_{435}$ data. Lines show the fraction of recovered over the implanted sources. The lines indicate results for simulated objects with different FWHM, ranging from 0.6 to 2.0 arcsec in steps of 0.2 arcsec, from right to left.

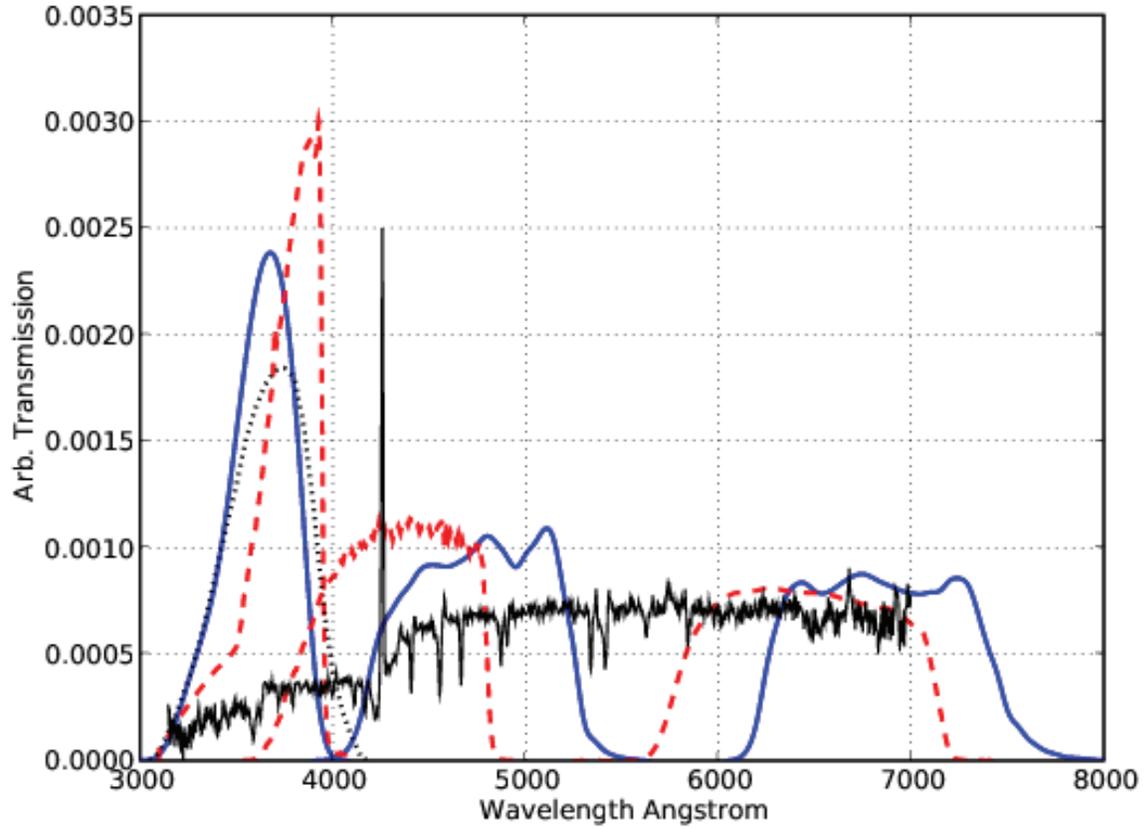

Fig. 17.— Comparison of the transmission (arbitrary units) between the Steidel $U_n$,G and R and the $U_V$, ACS $B_{435}$ and $R_V$ filters (dashed). The composite spectrum of an LBG at z=2.5 (Shapley et al. 2003) is also indicated.

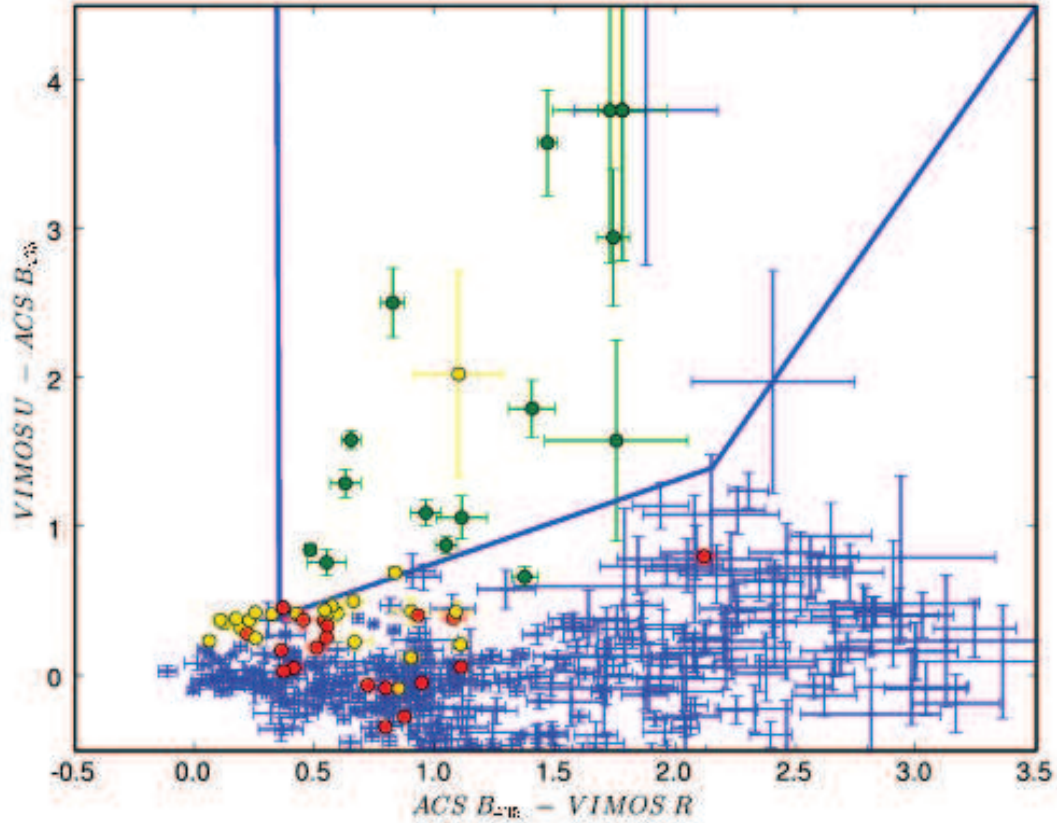

Fig. 18.— Definition of the Lyman break U-dropout color selection box in the ACS $B_{435}$ - $R_V$ versus $U_V$ - ACS $B_{435}$ plane. The green circles indicate galaxies with spectroscopic redshift $\geq 2.8$, yellow circles indicate galaxies with $2.0 \leq z \leq 2.8$, while red circles are galaxies with $1.8 \leq z \leq 2.0$. The blue symbols are foreground ($z \leq 1.8$) galaxies from FORS2 spectroscopic observations and with $R_V > 23.5$.

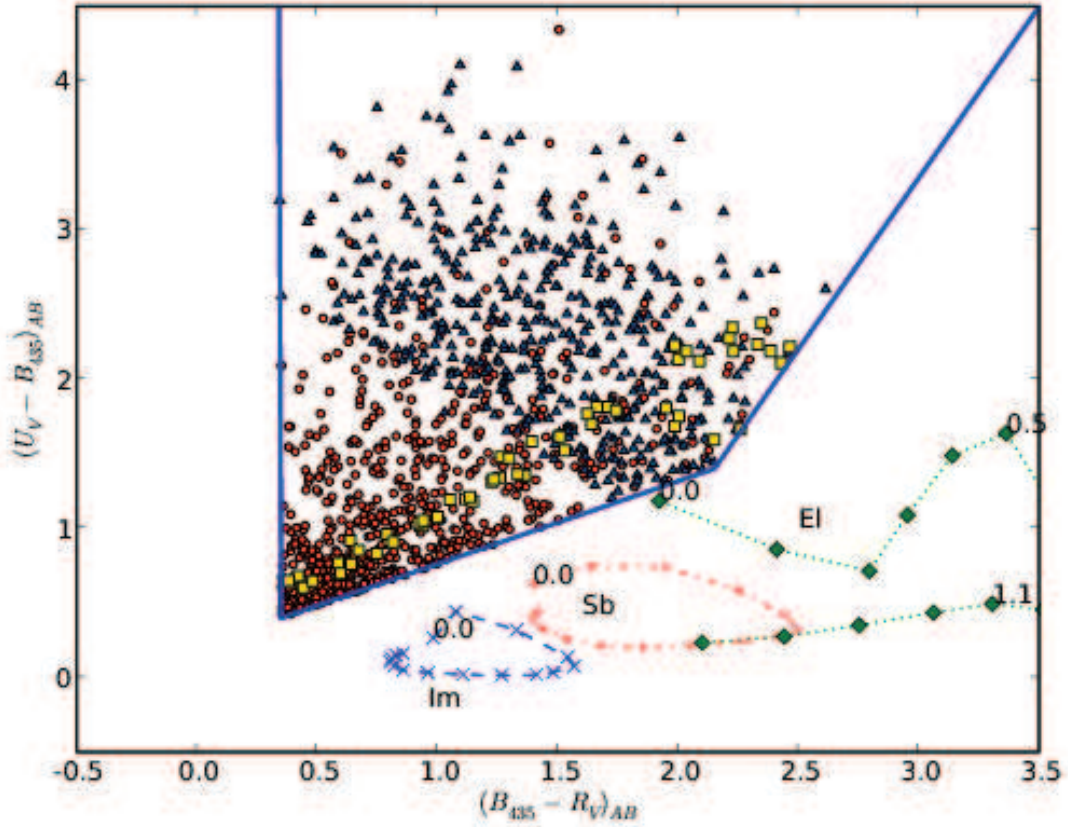

Fig. 19.— Lyman break selection based on $U_V$, ACS $B_{435}$ and $R_V$. The lines indicate the synthetic colors for an elliptical galaxy template from (Coleman et al. 1980) in the redshift range 0.0 to 1.1 (green), as well as Scd and Im templates in the redshift range 0.0 to 1.5 (red and blue respectively). Yellow squares are the expected colors of stars from the Pickles library (Pickles 1998) in the selection box. Red circles are the selected LBG with $U_V \leq 30.0$ while the blue triangles are the selected LBG with only limits for $U_V$ detection.

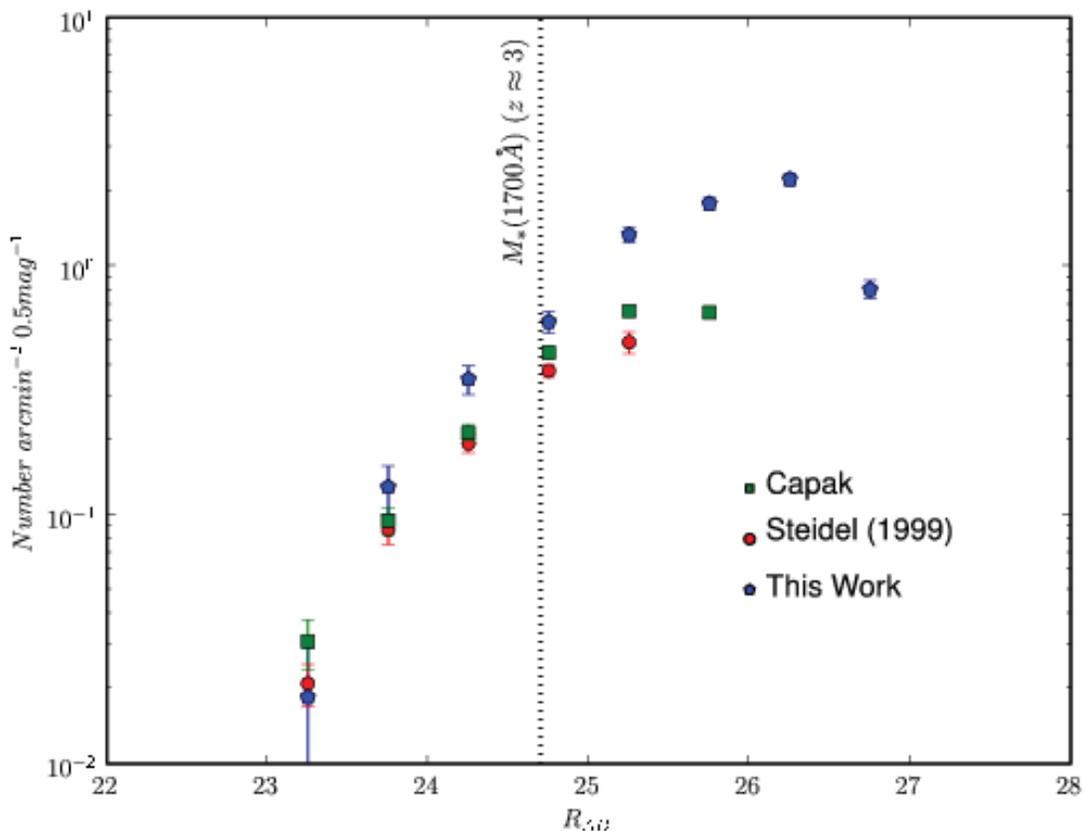

Fig. 20.— Raw number counts of U-dropout LBG, uncorrected for contamination or incompleteness, from the VIMOS–ACS GOODS-South data, compared to other measurements from the literature. The error bars in our data and in the raw counts from Capak et al. (2004) are $1\sigma$ Poisson fluctuations, while those in Steidel et al. (1999), corrected for interlopers contamination, include an estimate of cosmic variance.

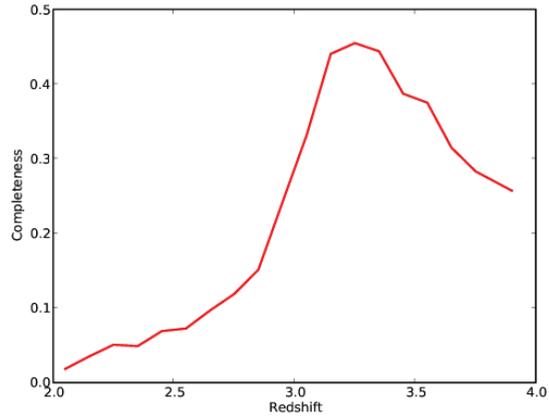

Fig. 21.— The expected redshift distribution of selected LBG, averaged over the magnitude range $23.5 < R < 27.5$, as determined from Monte Carlo simulations.

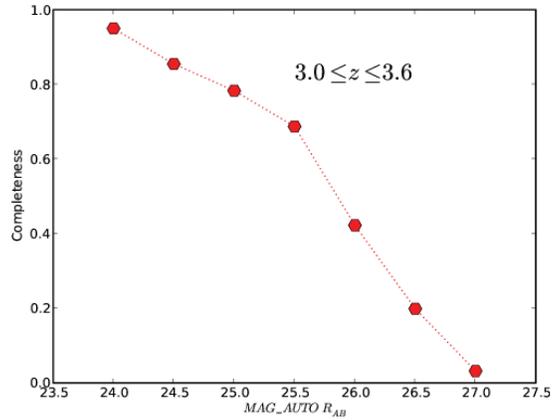

Fig. 22.— Completeness of LBG color selection versus R magnitude from Monte Carlo simulations, for the redshift interval $3.0 < z < 3.6$.

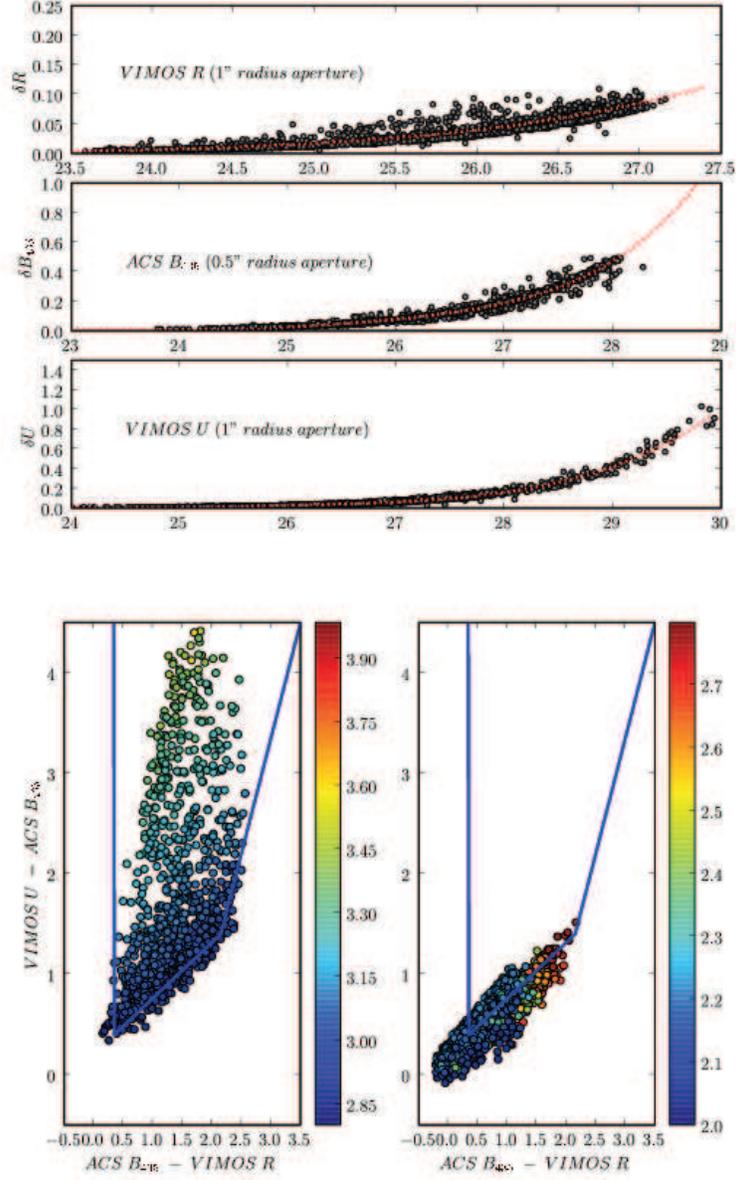

Fig. 23.— The effects of including photometric error in LBG selection. *Left:* The results of the fitting of the error in the aperture magnitude (see text). *Right:* Effect of the error: few $z \geq 2.8$ tend to leave the selection box (left) while lower redshift objects tend to enter it (right)

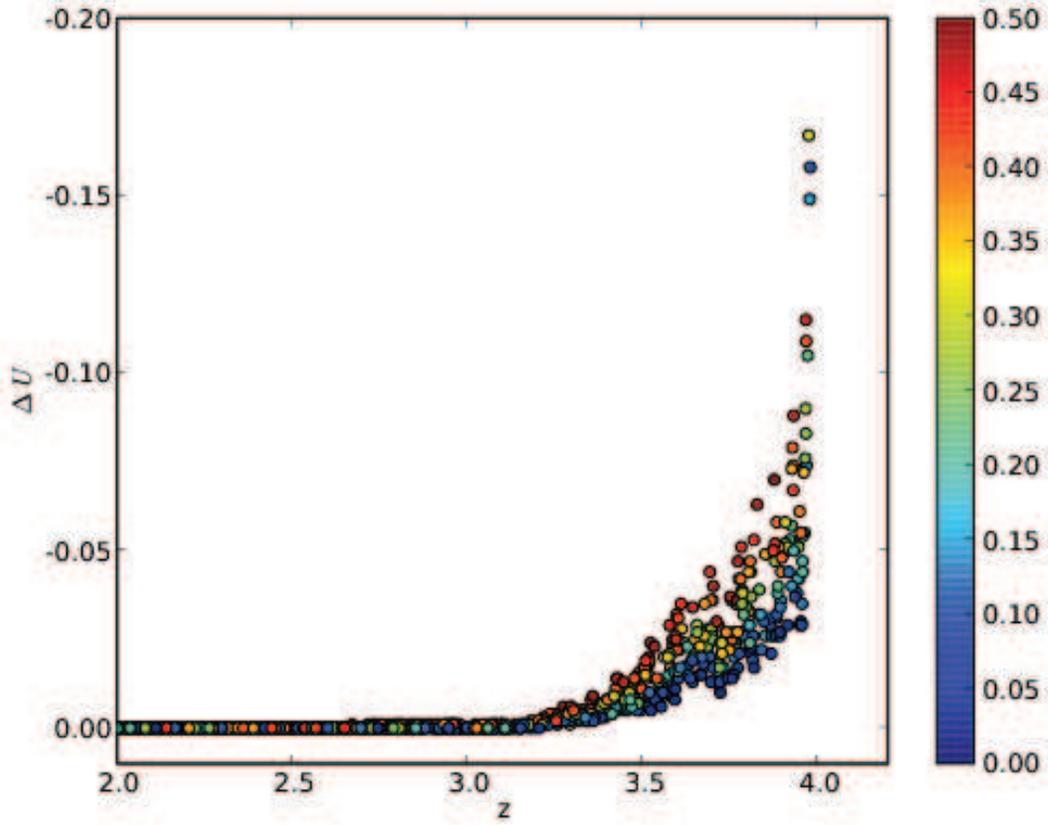

Fig. 24.— The effect of red leak in VIMOS quadrant 4 on LBG selection. The difference, $\Delta U = U_V(\text{red leak}) - U_V(\text{no red leak})$, is plotted against the redshift with internal $E(B-V)$ as color coding.

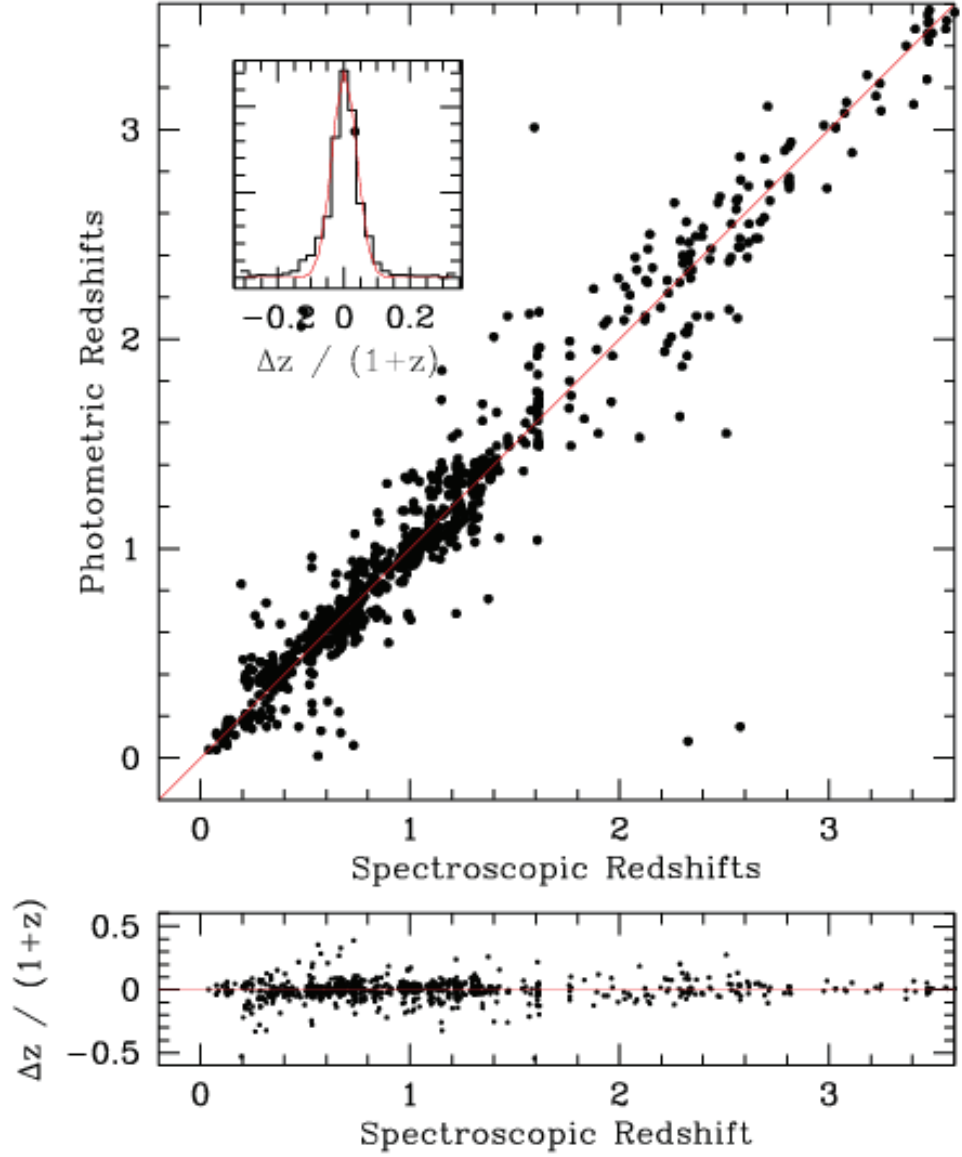

Fig. 25.— Photometric redshifts from the MUSIC catalog (see text) including the $U_V$ band ($\sigma_z = 0.085$).

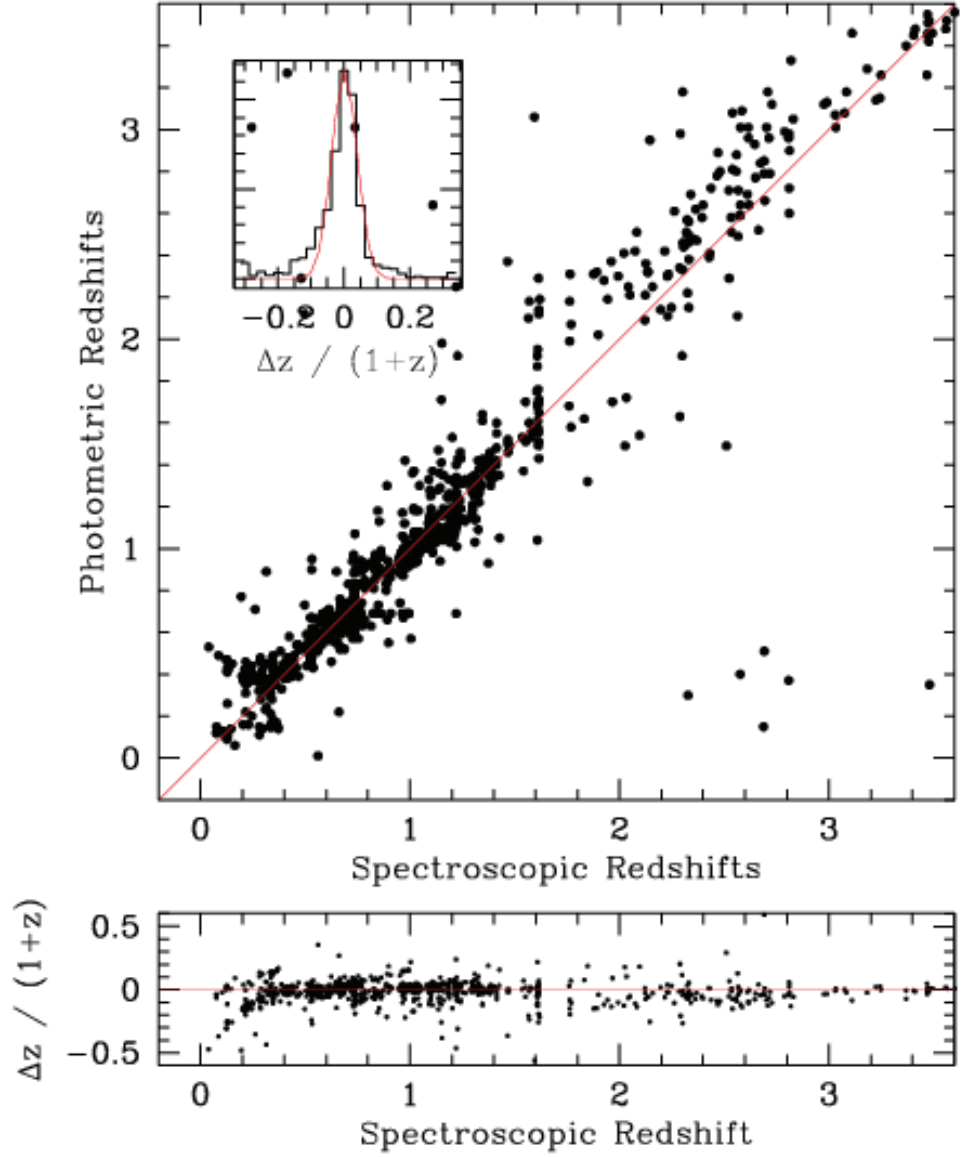

Fig. 26.— Photometric redshifts from the MUSIC catalog (see text) without the $U_V$ band ($\sigma_z = 0.094$).

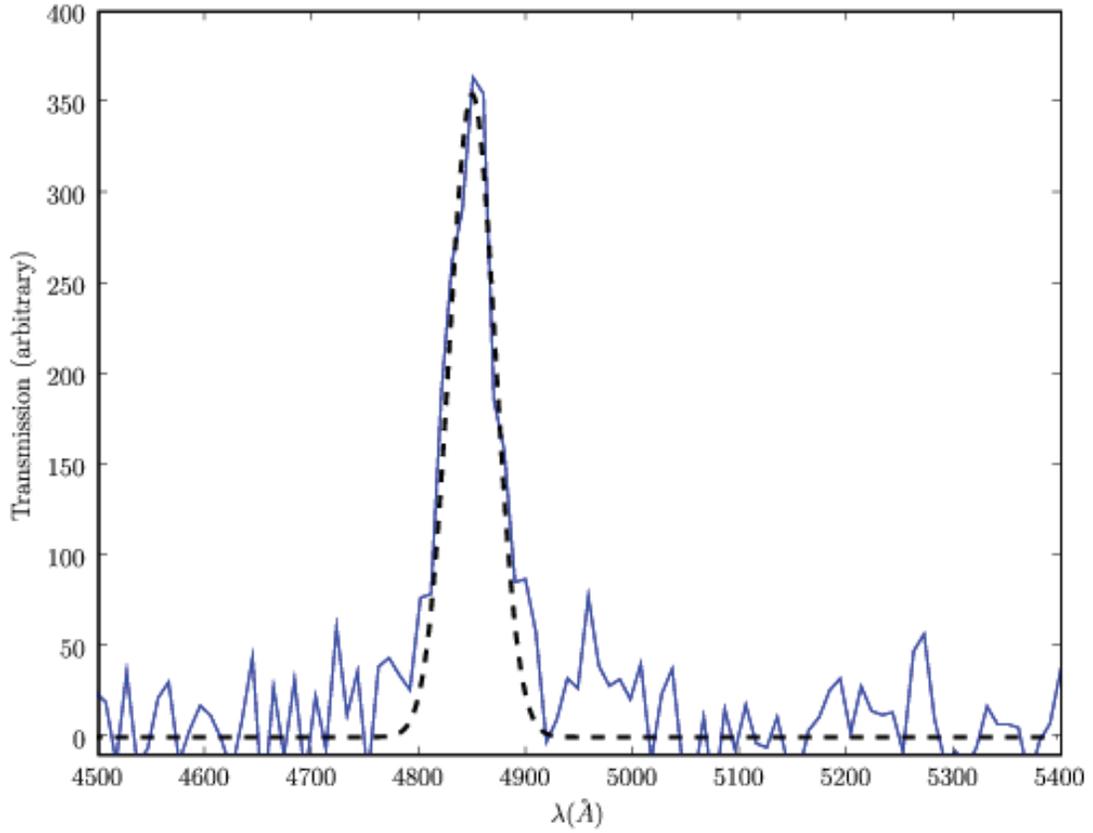

Fig. 27.— The red leak in the quadrant 4 of VIMOS camera, as found in the spectrum of *LTT 7379*. The gaussian fit is also shown (dashed line).